\documentclass[preprint,12pt]{aastex}

\newcommand{\kms}{\hbox{\,km s}^{-1}} 
\newcommand{\pc}{\hbox{\, pc}}
\newcommand{\half}{{\textstyle{1\over2}}}
\newcommand{\btheta}{\mbox{\boldmath $\theta$}}
\newcommand{\be}{\begin{equation}} 
\newcommand{\ee}{\end{equation}}
\newcommand{\yr}{\hbox{\,yr}}
\begin{document}

\title{Eccentric-disk models for the nucleus of M31}

\author{Hiranya V. Peiris\altaffilmark{1} and Scott
Tremaine\altaffilmark{2}} \affil{Department of Astrophysical Sciences,
Princeton University,  Princeton, NJ 08544}

\altaffiltext{1}{hiranya@astro.princeton.edu}
\altaffiltext{2}{tremaine@astro.princeton.edu}

\begin{abstract}
We construct dynamical models of the ``double'' nucleus of M31 in which the
nucleus consists of an eccentric disk of stars orbiting a central black
hole. The principal approximation in these models is that the disk stars
travel in a Kepler potential, i.e., we neglect the mass of the disk relative
to the black hole. We consider both ``aligned'' models, in which the eccentric
disk lies in the plane of the large-scale M31 disk, and ``non-aligned''
models, in which the orientation of the eccentric disk is fitted to the
data. Both types of model can reproduce the double structure and overall
morphology seen in Hubble Space Telescope photometry. In comparison with the
best available ground-based spectroscopy, the models reproduce the asymmetric
rotation curve, the peak height of the dispersion profile, and the qualitative
behavior of the Gauss-Hermite coefficients $h_3$ and $h_4$. Aligned models
fail to reproduce the observation that the surface brightness at P1 is higher
than at P2 and yield significantly poorer fits to the kinematics; thus we
favor non-aligned models. Eccentric-disk models fitted to ground-based
spectroscopy are used to predict the kinematics observed at much higher
resolution by the STIS instrument on the Hubble Space Telescope (Bender et
al.\ 2003), and we find generally satisfactory agreement.
\end{abstract} 



\section{Introduction} \label{intro}

The curious asymmetric nucleus of the Andromeda Galaxy (M31=NGC 224)
has intrigued astronomers since it was first resolved by the
balloon-borne telescope Stratoscope II in 1971 \citep{lds74}. Thus the
nucleus of M31 was a prime target for the Hubble Space Telescope
(HST). Early HST photometry (Lauer et al.\ 1993, hereafter L93; King, Stanford
\& Crane 1995) revealed that the apparent asymmetry arose because the
nucleus was double, with two components separated by $0\,\farcs5$; the
fainter peak P2 corresponds closely to the dynamical and photometric
center of the bulge, and the brighter peak P1 is off-center, with the
nucleus confined roughly within $2\arcsec$ of P2. Ground-based
spectroscopy, although unable to resolve the two components, revealed
a prominent velocity-dispersion peak and strong rotation-curve
gradients that implied the presence of a massive black hole
\citep{dr88,kor88,bac94,vdm94}.

It was natural to assume at first that P1 and P2 were orbiting star
clusters. However, the orbit of these clusters would decay by dynamical
friction from the bulge in $\lesssim 10^8\yr$, so the present configuration is
highly improbable. Difficulties with this and other hypotheses are
discussed by L93 and \citet[hereafter T95]{tre95}. 

A more attractive possibility, suggested by T95, is that the nucleus consists
of an eccentric disk of stars orbiting a massive black hole (hereafter
BH). In this model, the stars travel on eccentric orbits with approximately
aligned apsides; the BH is located at P2, and the bright off-center
source P1 is the apoapsis region of the eccentric disk. T95 argued that an
eccentric disk can reproduce most of the features seen in the HST photometry
and ground-based spectroscopy.

The model in T95 had several serious limitations. For example, it uses only
three Keplerian ringlets rather than a continuous distribution of orbits; the
eccentricity and inclination dispersion in the disk are modeled approximately
using a Gaussian point-spread function (PSF); the effect of the self-gravity
on the disk on the stellar orbits is neglected; and the model parameters are
chosen by eye, rather than by numerical parameter fitting.

The eccentric-disk model requires that the apsides of the disk stars precess
uniformly so that the disk maintains its apsidal alignment. The uniform
precession rate is simply the pattern speed of the eccentric disk. 
\citet{sta99} investigated the restrictions that this condition imposes on the
parameters of the eccentric disk. He argued that uniform precession requires
that the disk have a steep eccentricity gradient and a change in direction of
the eccentricity vector, in the sense that stars in the inner part of the disk
have apoapsides aligned on the P1 side, while stars in the outer part have
periapsides aligned on the P1 side. 

\citet{ss01} and \citet{sam02} have presented eccentric-disk models for the
nucleus of M31 in which the apsides of the disk stars precess uniformly. These
models are constructed by numerical integration of orbits in the combined
potential of a BH and a nuclear disk that is determined self-consistently from
the mass distribution of the orbits; in this respect their models are better
than the ones in this paper, which neglect the effect of the disk potential on
the orbits. However, both papers assumed that the disk is razor-thin, which is
unlikely to be correct---two-body relaxation alone will thicken the
disk to an axis ratio $\sim0.2$ if the disk age is comparable to the Hubble
time (see \S\ref{sec:relax}). $N$-body simulations of the disk by
\citet[hereafter B01]{bac00} were also mostly restricted to two dimensions;
some three-dimensional simulations were carried out but not described in
detail. The assumption of a razor-thin disk drove \citet{sam02} to models in
which the eccentric disk is not aligned with the larger M31 disk (we shall
confirm the claim of T95 that a thick eccentric disk aligned with the M31 disk
can provide a reasonably good fit to the data, but like Sambus \& Sridhar and
B01 we find that non-aligned models fit better than aligned
ones). \citet{ss01} did not have to consider a non-aligned disk because they
compared the photometry to the data only along the major axis.  Salow \&
Statler estimate a pattern speed of $14\kms\!\pc^{-1}$ while
\citet{sam02} find $16\kms\!\pc^{-1}$. The close agreement between these two
estimates does not imply that they are reliable, since the pattern speed makes
only a small contribution to the overall kinematics (at the P1--P2 separation a
pattern speed of $15\kms\!\pc^{-1}$ corresponds to a velocity $\lesssim 20\%$
of the peak rotation velocity), and since the dynamical models used in the two
papers are quite different (the mass ratio of the eccentric disk to the BH is
0.03 in Salow \& Statler and 0.65 in Sambhus \& Sridhar). 

B01 carry out $N$-body simulations of an eccentric disk that is
intended to resemble the nucleus of M31. Their experiments demonstrate that
long-lived eccentric configurations arise naturally from a variety of
non-axisymmetric (lopsided) initial conditions. Their best-fit simulation
reproduces the photometry and rotation curve of the M31 nucleus reasonably
well. However, once again they assume that the disk is thin (axis ratio
0.1). They estimate that the pattern speed is $\sim 3\kms\!\pc^{-1}$, far
smaller than the estimates of Salow \& Statler and Sambhus \& Sridhar.

A variety of new, high-resolution, space- and ground-based photometry and
spectroscopy of the nucleus of M31 has become available since 1995. These
include multicolor WFPC-2 images from HST with greater signal-to-noise ratio
(S/N), spatial resolution, and dynamic range than L93 \citep[hereafter
L98]{lau98}, which have been analyzed in detail by \citet{pen02};
near-infrared (JHK) photometry using both adaptive optics
\citep{dav97} and the NICMOS instrument on HST \citep{cor01}; ground-based
long-slit spectroscopy with slit width$=$0\,\farcs35 and seeing
FWHM$=$0\,\farcs64 \citep[hereafter KB99]{kb99}; spectroscopy with the Faint
Object Camera \citep[hereafter S99]{skcj99} and Faint Object Spectrograph of
HST \citep{tsv98}; integral-field spectroscopy using adaptive optics, with
FWHM$=$0\,\farcs5 (B01); and long-slit spectroscopy by the Space
Telescope Imaging Spectrograph (STIS) on HST (B01, Bender et al.\ 2003).

Given the wealth of new observations and the limitations of existing models,
the time is ripe to revisit the eccentric-disk hypothesis and confront more
realistic models with more accurate, varied and higher resolution data. In
\S\ref{obsintro} we review the rich phenomenology of the M31 nucleus, and in
\S\ref{t95intro} we briefly describe the eccentric-disk model. In
\S\ref{eccdiskmodel}, we construct an improved kinematic eccentric-disk model
with a range of free parameters that we can fit to the observations. In
\S\ref{nummeth} we describe the details of the numerical implementation of the
model and parameter fitting. We describe our results in \S\ref{results}, and
provide a discussion in \S\ref{discuss}.

We shall not attempt to fit or discuss all of the observations described
above. Instead, we shall focus on fitting the HST photometry described by L98
and the ground-based spectroscopy of KB99. As a test of our model, we shall
compare its predictions to HST spectroscopy from \citet{ben03} (see
B01 for an earlier analysis of the same data). 

Constructing general, self-consistent models of stellar systems to compare
with observations is a difficult task that is important in many areas of
galactic structure. The state of the art is orbit-based models of axisymmetric
systems, in which the distribution function can depend on up to three
integrals of motion \citep{cre99,geb00,ver02}. Eccentric-disk models are
substantially more complicated, because they are non-axisymmetric and involve
up to five integrals of motion. To keep the problem manageable we shall adopt
a number of simplifying approximations. Of these, the most important is that
the stars follow Kepler orbits in the gravitational field of the BH; that is,
we neglect the effect of the self-gravity of the eccentric disk on the stellar
orbits. The effects of this approximation are discussed briefly in
\S\ref{discuss}.

Throughout this paper, we adopt KB99's estimates of the distance and
foreground extinction to M31, $d=0.77$ Mpc and $A_V=0.24$ \citep{bh84}. At this
distance $1\arcsec=3.73\pc$.

\subsection{Observations} \label{obsintro}

The observational facts can be summarized as follows:

\begin{enumerate}

\item The nucleus is composed of two components, separated by
0\,\farcs49 and labeled P1 and P2. The two components cannot be decomposed
into a superposition of two systems with elliptical isophotes (L93). The total
bulge-subtracted magnitude of both components of the nucleus is $V=12.55\pm
0.2$, corresponding to a total luminosity $L=6.0\times10^6L_\odot$
(KB99). The same double structure is seen in near-infrared images,
demonstrating that the apparent duplicity of the nucleus is not due to a dust
band \citep{dav97,cor01}. There is no evidence that the photometry is
affected by spatially varying obscuration.

\item The position angle of the line joining P1 and P2 is
$43\arcdeg\pm1\arcdeg$ (as usual, measured eastward from north), while the
position angle of the isophotes within 1\,\farcs0 of the center of P1 is about
$63\arcdeg$ (L93); thus there is an isophote twist of about $20\arcdeg$. The
position angle of the outer parts of the nucleus (1--$2\arcsec$) is
$55\arcdeg\pm1\arcdeg$, and the position angle of the large-scale M31 disk is
$38\arcdeg$.

\item A compact blue source is centered on P2 (we shall call this
P2B); as a result of this source, P2 is actually brighter than P1 in both the
near- and far-UV (L98; King et al.\ 1995). King et al.\ (1995) suggest that
P2B is a low-level AGN; however, the higher S/N and resolution of L98's data
reveal that P2B is extended, with a half-power radius 0\,\farcs2 and position
angle $62\arcdeg\pm8\arcdeg$ (B01). STIS spectra reveal that P2B is a cluster
of early-type stars, with the remarkably high velocity dispersion of
800--$950\kms$ \citep{kor02,ben03}.

\item The stellar component of P2, which surrounds P2B, exhibits a
weak cusp, $I(r)\propto r^{-\gamma}$ with $\gamma\simeq0.1$ (L93). The
peak in the stellar component appears to be slightly offset from P2B; the
measured offset is strongly wavelength-dependent because of the color
difference. 

\item The central surface brightness of P1 is $\mu_V=13.4\hbox{ mag
arcsec}^{-2}$ (as measured in a slit 0\,\farcs22 wide; see L93) and the
major-axis core radius is about 0\,\farcs4. In contrast to P2, P1 exhibits no
compact blue source and no cusp in the starlight (L93; King et al.\ 1995); the
central surface brightness of P2 in the same slit is $13.7\hbox{ mag
arcsec}^{-2}$, 0.3 mag fainter.

\item The source P2 is very close to the center of the galaxy (within
0\,\farcs1 according to L98) as defined by the centroids of the bulge
isophotes just outside the nucleus; KB99 estimate that P2B is displaced from
the bulge center by 0\,\farcs07 in the anti-P1 direction. 

\item The $V-I$ color of the nucleus appears to differ from the surrounding
bulge (although L98 and B01 disagree on the color difference); the
color difference implies that the stellar populations in the bulge and nucleus
are different, while the populations in P1 and P2 (outside P2B) are the same
(L98). KB99 reach a similar conclusion from absorption-line strengths.

\item The gradients in both the rotation curve and the
velocity-dispersion profile are unresolved or marginally resolved, that is,
they become steeper as the resolution improves. Both profiles are
asymmetric. KB99 find that the maximum (bulge-subtracted) rotation speed is
$-236\pm4\kms$ on the anti-P1 side but only $179\pm2\kms$ on the P1 side (see
Table \ref{tab:kinres}). S99 claim that they have resolved the central
gradient in the rotation curve, at $1000\kms\hbox{ arcsec}^{-1}$; the rotation
curves from S99 and KB99 are consistent when S99's HST results are smoothed to
KB99's ground-based resolution. The peak velocity dispersion is displaced from
P2 in the anti-P1 direction, by $0\,\farcs06 \pm 0\,\farcs03$ according to S99
(without bulge subtraction) or 0\,\farcs13 according to KB99 (with bulge
subtraction). The peak dispersion is $287 \pm 9\kms$ according to KB99 (with
bulge subtraction), and $440\pm 70\kms$ according to S99 (without bulge
subtraction). With bulge subtraction the peak dispersion in the S99 data would
be even higher, approaching a factor of two larger than KB99. Some of this
difference presumably reflects the higher resolution of HST but it is also
possible that the S99 dispersions are systematically high.
Remarkably, at 1\arcsec\ away from the nucleus (in either direction along the
major axis) the dispersion has fallen by a factor of three or more, to
$\lesssim100\kms$. B01 argue that S99 have made a 0\,\farcs074 error
in registering the KB99 spectroscopic data with their HST data. 

\item The zero-point of the rotation curve (relative to the systemic
velocity of the bulge) is displaced from P2B toward P1. KB99 find that the
displacement is $0\,\farcs051\pm 0\,\farcs014$ S99 find
$0\,\farcs13\pm0\,\farcs05$ (composed of $0\,\farcs16\pm0\,\farcs05$ offset
from the stellar peak of P2 minus 0\,\farcs025 offset between the stellar peak
of P2 and P2B) and B01 find 0\,\farcs 031. These differences may reflect the
small offset between P2B and the stellar peak of P2, which means that the
measured position of the peak brightness of P2B is wavelength-dependent, as
well as differences in spatial resolution (see B01 for discussion).

\item KB99 also measure the Gauss-Hermite coefficients of the
line-of-sight velocity distribution. They find that the zero-point of
$h_3$ is displaced by about 0\,\farcs04 from the zero-point of the
rotation curve, in the anti-P1 direction.

\end{enumerate}

\subsection{The eccentric-disk model} \label{t95intro}

In this model, the nucleus contains a single BH located at the center of
P2B. This is consistent with (i) the presence of a compact cluster of young
stars, perhaps formed from a gas disk surrounding the BH (the stellar
collision rate is too slow to explain the presence of these stars; see Yu
2003); (ii) the high velocity dispersion of the P2B cluster; (iii) the weak
surface-brightness cusp observed in P2 outside P2B; (iv) the presence of an
unresolved dispersion peak in the old stars close to P2 but outside P2B.

Most or all of the stars in the nucleus are assumed to lie in a disk
surrounding the BH. The mass of the stellar disk is assumed to be small
compared to the mass of the BH (see \S\ref{discuss} for a discussion of this
approximation). Thus, the stars travel on approximately Keplerian orbits, and
the asymmetry of the nucleus arises because the orbits are eccentric and the
mean eccentricity vector is non-zero (i.e. the apsides are approximately
aligned). The off-center source P1 marks the apoapsis region of the disk,
which is bright because stars linger at apoapsis, while P2 (outside P2B)
represents a combination of disk stars with smaller semimajor axes and stars
with periapsis near P2 and apoapsis near P1.

The disk is assumed to be in a steady state (there is presumably slow figure
rotation as the apsides precess due to the self-gravity of the disk and the
tidal field from the bulge, but we neglect the effect of figure rotation on
the kinematics). The center of mass of the stellar disk plus the BH must
therefore lie at the center of the bulge. It is natural to assume that the age
of the disk is comparable to the age of the galaxy, i.e. $\sim10^{10}$ yr,
although this is not required for the model.

The eccentric-disk model is consistent with the color and line-strength
observations, which imply that the stellar populations in the bulge and
nucleus are different, while the populations in P1 and P2 (outside P2B) are
the same. It is also consistent with the observation that the BH at P2 is
located close to the center of the bulge, which is expected if the BH mass is
much larger than the mass of the stars in the eccentric disk. Alternative
models in which both both P1 and P2 are stellar clusters bound to BHs would
suggest that both should have a cusp and perhaps a compact blue source, but P1
has neither.

It is natural to assume at first that the eccentric nuclear disk lies in the
plane of the large-scale M31 disk (``aligned models''), although we shall find
that ``non-aligned'' models in which this is not the case provide better fits
to the data. In aligned models the disk must be relatively thick: M31 is
highly inclined (inclination $77\arcdeg$), so the disk is seen nearly edge-on
and the isophotes of a thin disk would be flattened too strongly.

The T95 model, although not unique, correctly predicted several
features that were discovered in subsequent observations, such as the
displacement of the zero-point of the rotation curve from P2B toward
P1, and the shape and amplitude of the asymmetry in the rotation
curve. S99 explored a number of similar eccentric-disk models and were
able to improve the kinematic fit, but found it difficult to reproduce
simultaneously the location of the zero-point of the rotation curve,
the steepness of the rotation curve, and the overall photometric
symmetry of the nucleus about P2 beyond the distance of P1 (see also the
discussion of S99's results in \S\S2.3 and 5.3 of B01). 

In order to gain some intuition for the features of a eccentric-disk
model, we review here the simple case of an infinitesimally thin, cold disk
with a BH at the origin \citep{sta99}. Stars are assumed to orbit on confocal,
nested ellipses. It is straightforward to show,
using standard formulae for Kepler orbits, that the surface density along the
$x$-axis is given by
\be 
\label{thindisk} 
\Sigma \left(a\right) = \frac{\mu(a)}{2\pi a}
\frac{1\pm e}{\sqrt{1-e^2}(1\pm e\pm ae')},  
\ee
where the $+$ or $-$ sign is chosen at apoapsis or periapsis respectively, 
$\mu(a)\equiv dm/da$, $dm$ is the mass contained
between $a$ and $a+da$, and $e'=de/da$.

Denoting the surface brightness at periapsis and apoapsis as $\Sigma_p(a)$ and
$\Sigma_a(a)$ respectively, the brightness ratio at a given value of $a$ is
\be 
\label{thinratio} 
\frac{\Sigma_a(a)}{\Sigma_p(a)} =
\frac{1-ae'/(1-e)}{1+ae'/(1+e)}. 
\ee
To meet the condition that the disk have a substantially higher surface
brightness at apoapsis we require $e' < 0$ and either $e$ or $|ae'|$
of order unity. Therefore, as already noted by T95 and S99, successful models
require a strong negative eccentricity gradient in the interval of semimajor
axes corresponding to P1 (see Fig.\ \ref{fig:meane}).

\section{A model for an eccentric disk} \label{eccdiskmodel}

\subsection{Coordinate systems}

We start by establishing our notation and defining the coordinate
systems that we will employ. We use the BH as the origin of all of our
coordinate systems, and assume that the BH coincides with the center
of the blue source P2B.

Our first coordinate system, the ``sky-plane'' system, is denoted by
$(X,Y,Z)$. The $(X,Y)$ plane is the sky plane; the positive $X$-axis points
west, the positive $Y$-axis points north, and the positive $Z$-axis points
along the line of sight toward the observer. In this system the line-of-sight
velocity is $V_{\rm los}=-\dot Z$; the minus sign is necessary for consistency
with the usual convention that objects receding from the observer have
positive velocity. Our second coordinate system is the ``disk-plane'' system,
denoted by $(x,y,z)$. The $(x,y)$ plane is defined to be the symmetry plane of
the eccentric nuclear disk. The $x$-axis is aligned with the major axis of the
nuclear disk (more precisely, the distribution of eccentricity vectors
[defined at the end of \S\ref{sec:orbel}] of the disk stars is assumed to be
symmetric about the $x$-axis).  A third ``orbital-plane'' coordinate system
$(x',y',z')$, different for each star, is defined so that the $(x',y')$ plane
is the orbital plane of the star, the positive $x'$-axis points toward the
periapsis of the star, and the positive $z'$-axis is parallel to the star's
orbital angular-momentum vector. All three coordinate systems are right
handed.

The transformation between the orbital-plane and disk-plane
coordinates is given by: \be \label{orb-disk} \left( \begin{array}{c}
x \\ y \\ z \end{array} \right) =  \left (
\begin{array}{ccc} \cos \Omega & -\sin \Omega & 0 \\ \sin \Omega &
\cos \Omega & 0 \\ 0 & 0 & 1 \end{array} \right) \left (
\begin{array}{ccc} 1 & 0 & 0 \\ 0 & \cos I & -\sin I \\ 0 & \sin I &
\cos I \\ \end{array} \right) \left( \begin{array}{ccc} \cos \omega &
-\sin \omega & 0 \\ \sin \omega & \cos \omega & 0 \\ 0 & 0 & 1
\end{array} \right) \left( \begin{array}{c} x' \\ y' \\ z' \end{array}
\right), \ee where $I$ is the inclination of the orbit relative to the
disk plane, $\Omega$ is the longitude of the ascending node of the
orbit on the disk plane (the angle in the disk equatorial plane from
the $x$-axis to the ascending node, where the orbit has $z=0$, $\dot
z>0$), and $\omega$ is the argument of periapsis (the angle in the
orbital plane from the ascending node to the $x'$ axis).

The transformation between the disk and sky-plane coordinates is given
by: \be \label{disk-sky} \left( \begin{array}{c} X \\ Y \\ Z
\end{array} \right) =  \left (
\begin{array}{ccc} \cos \theta_l & -\sin \theta_l & 0 \\ \sin \theta_l
& \cos \theta_l & 0 \\ 0 & 0 & 1 \end{array} \right) \left (
\begin{array}{ccc} 1 & 0 & 0 \\ 0 & \cos \theta_i & -\sin \theta_i \\
0 & \sin \theta_i & \cos \theta_i \\ \end{array} \right) \left (
\begin{array}{ccc} \cos \theta_a & -\sin \theta_a & 0 \\ \sin \theta_a
& \cos \theta_a & 0 \\ 0 & 0 & 1 \end{array} \right) \left (
\begin{array}{c} x \\ y \\ z \end{array} \right).
\ee Here, $\theta_l$ is the angle in the sky plane from the $X$-axis
to the ascending node of the disk on the sky (i.e. the point where a
star orbiting in the disk plane has $Z=0$ and $\dot Z>0$). The
inclination angle $\theta_i$ is the angle between the normals to the
sky plane and the disk plane ($\cos\theta_i=\hat{\bf z}\cdot\hat{\bf
Z}$).  The azimuthal angle $\theta_a$ is measured in the disk plane
from the ascending node of the disk on the sky to the symmetry axis of
the disk (the positive $x$-axis). All angles are measured in the usual
sense, counterclockwise as seen from the positive $Z$, $z$, or $z'$
axis.

If the symmetry plane of the disk coincides with the symmetry plane of the
large-scale M31 disk, then the parameters $\theta_l$ and $\theta_i$ are
determined by the orientation of M31. The apparent major axis of the M31 disk
has position angle $\hbox{PA}_{\rm gal}=37\fdg 7$ \citep{dev58}, and
$\theta_l=\hbox{PA}_{\rm gal}\pm\half\pi$. The SE side of M31 is approaching,
and the near side of M31 is to the NW \citep{hod92}, so the correct choice is
$\theta_l=\hbox{PA}_{\rm gal}-\half\pi=-52\fdg 3$.  The inclination angle
$\theta_i=77\fdg 5$ \citep{hod92}.

\subsection{Orbits}\label{sec:orbel}

In the models described here, we shall neglect the gravitational
influence of the bulge and the nuclear disk on the disk-star orbits;
that is, we assume that the disk stars travel on Keplerian orbits
under the influence of a central BH of mass $M_\bullet$. This approximation is
discussed more fully in \S\ref{discuss}.

The orbital period of a star traveling in a Keplerian orbit around the BH is
$P=2\pi/n$, where $n=(\mu/a^3)^{1/2}$, $\mu=GM_\bullet$ and $a$ is the
semimajor axis.  Its coordinates $(x',y',z'=0)$ are given parametrically by
the relations 
\be 
\label{orb-x-coord} 
x'=a(\cos E-e), \qquad y'=a\sqrt{1-e^2}\sin E, \qquad M =E-e\sin E, 
\ee 
where $e$ is the eccentricity, $E$ is the eccentric anomaly, and $M$ is the
mean anomaly, defined as $2\pi t/P$ where $t$ is the time since periapsis
passage.

The velocity components of the star are given by \be \label{orb-x-vel}
\dot{x'} = -na\frac{\sin E}{1- e\cos E}, \qquad  \dot{y'} =
na\frac{\sqrt{1-e^2}\cos E}{1- e\cos E}, \qquad \dot z'=0.  \ee The
corresponding line-of-sight velocity $V_{\rm los}$ is then obtained by
differentiating equations (\ref{orb-disk}) and (\ref{disk-sky}) with
respect to time.

We shall also use the canonical Delaunay variables (e.g.\ Dermott \&
Murray 1999),

\parbox{0.8cm}{
\begin{eqnarray*} 
\theta_1 &=& M, \\ \theta_2 &=& \omega, \\ \theta_3 &=& \Omega,
\end{eqnarray*}}
\hfill\parbox{15cm}{
\begin{eqnarray} 
J_1 &=& \sqrt{\mu a}, \\ J_2 &=& \sqrt{\mu a (1-e^2)}, \\ J_3 &=&
\sqrt{\mu a (1-e^2)} \cos I.
\label{eq:delaunay}
\end{eqnarray}}

The longitude of periapsis is $\varpi=\Omega+\omega$. It is also
convenient to introduce the eccentricity vector ${\bf e}$, with
components in the $(x,y,z)$ coordinate system
$(e\cos\varpi,e\sin\varpi,0)$.

\subsection{The distribution function}

The nuclear disk is described by its distribution function $f({\bf x},{\bf
v})$, defined so that $f({\bf x},{\bf v})d{\bf x}\,d{\bf v}$ is the mass or
light contained in the phase-space volume element $d{\bf x}\,d{\bf
v}$. The corresponding volume element in orbital elements is
$d{\btheta}d{\bf J}=\half\mu^{3/2}a^{1/2}e\sin I\,da\,de\,dI$. 
According to Jeans's theorem, the distribution function (hereafter DF)
can only depend on the integrals of motion, which in a Kepler potential can be
taken to be $a$, ${\bf e}$, $I$, $\Omega$. Since the DF is a function of five
variables (only two of the components of ${\bf e}$ are independent, since
${\bf e}$ lies in the $z=0$ plane), and we can only measure three (the
sky-plane coordinates $X$ and $Y$, and the line-of-sight velocity $V_{\rm
los}$), we cannot, even in principle, completely determine the DF directly
from the observations. Instead, we choose a plausible parametric form for the
DF, whose parameters are then fit to the observations. An alternative
non-parametric approach would be to find the maximum-entropy DF that is
consistent with the observations.

Thus we assume that 
\be 
\label{df} 
f(a,{\bf e},I)=g(a)
\exp\left[-\frac{[{\bf e}-{\bf e}_m(a)]^2}{2\sigma_e^2}\right]
\exp\left[-\frac{I^2}{2\sigma_I(a)^2}\right], 
\ee 
where ${\bf e}_m(a)\equiv e_m(a)\hat{\bf x}$ is the mean eccentricity vector
(note that $e_m$ can have either sign). The magnitude of the mean eccentricity
vector depends on semimajor axis but its direction is assumed to be fixed;
this is consistent with the observation that gas-free stellar disks generally
do not exhibit spiral structure, and with the anti-spiral theorem
\citep{lbo67}. Note that $\sigma_e$ is the
dispersion in one component of the eccentricity vector, not the dispersion in
eccentricity; thus, for ${\bf e}_m=0$, the rms eccentricity is
$\sqrt{2}\sigma_e$. Similarly, for small inclinations the rms inclination is
$\sqrt{2}\sigma_I$. For small eccentricity and inclination, the rms thickness
at a given radius is simply $\langle z^2\rangle^{1/2}=a\sigma_I(a)$. 

For simplicity, the width of the eccentricity distribution, $\sigma_e$, is
assumed to be independent of semimajor axis. However, if the disk is aligned
with the plane of M31 and the width of the inclination distribution,
$\sigma_I(a)$, is independent of semimajor axis (so that the disk thickness is
proportional to radius), the models are too flattened near P2. Thus we choose
the form
\be
\label{isigmaform} 
\sigma_I(a) = \sigma_I^0 \exp\left(-a/a_{I}\right).
\ee
\noindent
This variation of $\sigma_I(a)$ with semimajor axis $a$ is much less important
for models that are not aligned with the plane of M31. In such models we have
found that the best-fit value of $a_I$ tends to be large enough that the
dependence of $\sigma_I(a)$ on semimajor axis is unimportant within the
nucleus.

Our numerical experiments show that the dispersion in
inclination $\sigma_I^0$ has a profound effect on the photometric
appearance of the disk at radii $\lesssim 1\arcsec$. First consider aligned
models. If the disk is too thin, the isophotes are too
flattened. If the disk is too thick, the surface brightness at P1 is
too low (i.e. smaller than at P2), and the ``valley'' between the
peaks in the surface brightness distribution gets filled up. In the
non-aligned model, if the disk is too thin, the edges of P1 appear too
well-defined, and the P1 contours are crescent-shaped, in contrast to the
more diffuse blob-like appearance of P1 in the HST photometry; also,
there is essentially no P2 peak, again in contradiction to the
data. If the disk is too thick, the effect on the photometry of a
non-aligned disk is the same as for an aligned disk. The effect of
$\sigma_I^0$ on the kinematics is less dramatic, but still important:
in the aligned model, for larger values of $\sigma_I^0$, the peaks of
the rotation curve are further apart, the maximum rotation speed is
higher, the rotation curve is more symmetric, and the displacement of
the dispersion peak from P2B is smaller. In the non-aligned model, if
the disk is too thin, the velocity extrema are too large compared to
the data, and the dispersion peak is too narrow; if it is too thick,
the converse happens.

We experimented with many forms for the mean eccentricity distribution
$e_m(a)$, including Gaussian, linear, and exponential forms. If the
eccentricity peaks at the origin, we found that the valley in the surface
brightness profile between P1 and P2 is not deep enough, so that P1 and P2 do
not appear as distinct as they do in the data. Thus, we concluded that the
mean eccentricity should peak away from the origin. As suggested in the
discussion following equation (\ref{thinratio}), we also found that
satisfactory models required that the eccentricity decline rapidly to near
zero in the semimajor axis interval from about 0\,\farcs5 to 1\,\farcs0.
We eventually chose the functional form
\be 
\label{eccform} 
e_m(a) = \alpha (a_e - a)\exp\left[-\frac{(a-a_g)^2}{2 w^2}\right].
\ee

\begin{figure}
\plotone{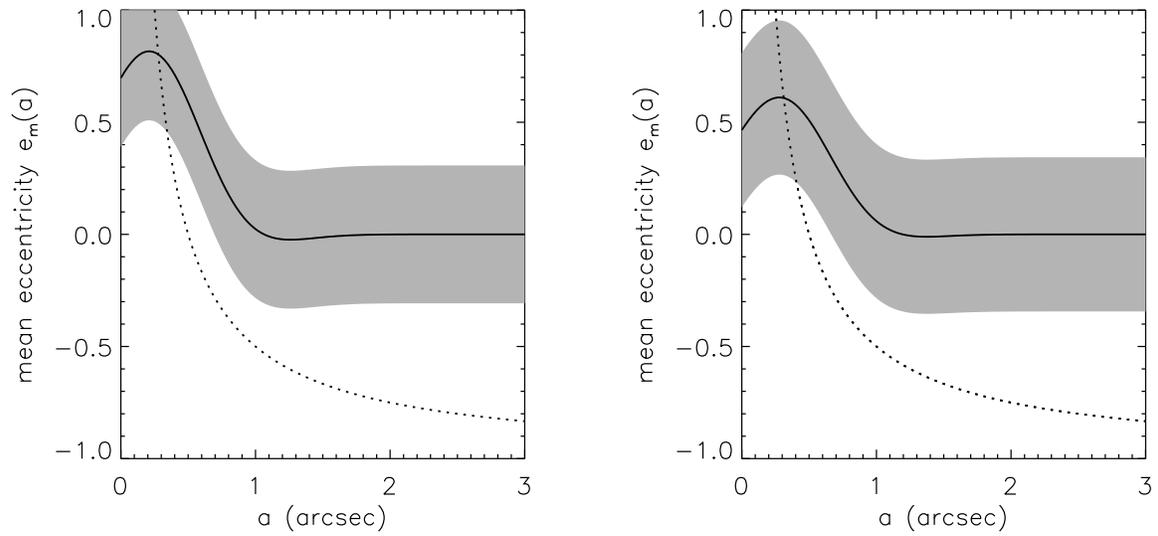}
\caption{The mean eccentricity $e_m(a)$ as a function of semimajor axis,
for the aligned (left) and non-aligned (right) models. The gray bands
represent 1--$\sigma$ deviations (parameter $\sigma_e$). Note that Monte-Carlo
points with $|e|>1$ are discarded. The dashed line is the locus of orbits with
apoapsis equal to 0\,\farcs5, the separation of P1 and P2.}
\label{fig:meane}
\end{figure}

The functional form (\ref{df}) that we have chosen for the DF allows
eccentricities that exceed unity, which are unphysical; thus, in choosing
orbital elements in our Monte Carlo simulation we discard points with $e\ge1$.
Figure \ref{fig:meane} shows $e_m$ for the best-fit aligned and non-aligned
models, along with the 1--$\sigma$ deviations represented by
$\sigma_e$. A similar figure is shown by B01 (Fig. 22); they find a similar
peak eccentricity $\simeq 0.7$, although at a slightly larger radius ($\simeq
0\,\farcs5$ instead of 0\,\farcs3), and a similar sharp decline in eccentricity
outside the peak.

To choose the function $g(a)$ that specifies the semimajor axis
distribution, we first convert the DF (\ref{df}) into an effective
surface density distribution; here the effective surface density
$\Sigma(a)$ is defined so that $2\pi a\Sigma(a)da$ is the mass with
semimajor axes in the range $[a,a+da]$. We have 
\be
\label{phase-elem} 
\Sigma(a_0) = {1\over 2\pi a_0}\int d{\bf
J}d{\btheta}\,\delta(a-a_0)f({\bf J},{\btheta}),  
\ee 
where the action-angle variables $({\bf J},\btheta)$ are defined in equation
(\ref{eq:delaunay}). For the DF (\ref{df}) we have
\be
\Sigma(a_0)= {\pi\mu^{3/2}\over a_0^{1/2}}g(a_0)\int_0^\pi dI\,\sin I
\exp\left[-\frac{I^2}{2\sigma_I(a_0)^2}\right]
\int_0^{2\pi} d\varpi \int_0^\infty de \, e \exp\left[-\frac{[{\bf
e}-{\bf e}_m(a_0)]^2} {2\sigma_e^2}\right].
\ee
The innermost two integrals are easily evaluated by writing
$e\,de\,d\varpi =d{\bf e}$ and using Cartesian coordinates for ${\bf
e}$: 
\be 
\label{annulus} 
\Sigma(a_0)= {2\pi^2\mu^{3/2}\sigma_e^2\over
a_0^{1/2}} g(a_0) \int_0^\pi dI\,\sin
I\exp\left[-\frac{I^2}{2\sigma_I(a_0)^2}\right].  
\ee  
Since $\sigma_I(a_0)$ is generally small, to a good approximation we have
\be 
\label{annulusa} 
\Sigma(a_0)=2\pi^2\mu^{3/2}\sigma_e^2\sigma_I^2(a_0)a_0^{-1/2}g(a_0).  
\ee

Initially, we tried a simple exponential for the effective surface-density
distribution, but we found that this form results in too much light at the
center, and too much contamination of the bulge by the disk at large
radii. After some experimentation, we concluded that to match the photometry
we require an effective surface-density distribution that has a central
minimum, an outer cut-off, and a maximum roughly at the P1 position. The
steepness of the inner gradient of the distribution is important. For example,
we experimented with a Gaussian distribution,
$\Sigma(a)=\Sigma_0\exp[-k(a-a_0)^2]$, and an exponential model with a central
hole, $\Sigma(a) = \Sigma_0 [\exp(-a/a_0)- \exp(-a/a_1)]$. We found that for
the former distribution the
best-fit model (viewed from above the disk) has a wider central hole with a
more well-defined boundary, a more well-defined P1 peak which extends for a
greater distance around the circumference of the disk, and less
surface brightness coming from the outer extremities of the disk.
When projected onto the sky plane, the
Gaussian model is less successful in reproducing the dip in the surface
brightness profile between the P1 and P2 peaks. The fact that the
double-exponential model is asymmetric about its maximum also seems to help in
reproducing the photometric data. In the case of the double-exponential model,
we consistently found best-fit models where $a_1
\sim a_0$, leading to an inner gradient rising as $a$. We also experimented
with models with $\Sigma(a) = \Sigma_0 a^n \exp(-a/a_0)$, and settled on $n=2$
as the most promising value of this parameter. Therefore, we choose the
following distribution, with a hole in the middle and an outer Fermi-type
cutoff,
\be 
\label{surfden} 
\Sigma(a) = \Sigma_0 \frac{a^2 \exp(-a/a_0)}{1+\exp\left[c_1(a-c_2)\right]};
\ee 
the fit is quite insensitive to the value of the cut-off parameter $c_1$. We
then define $g(a)$ using equation (\ref{annulusa}).

In total, our model for the DF depends on 10 parameters: two for the
semimajor axis distribution ($a_0$, $c_1$, $c_2$); two for the
inclination distribution ($\sigma_I^0$, $a_I$), and five for the
eccentricity distribution ($\alpha$, $a_g$, $a_e$, $w$, $\sigma_e$).

\subsection{The bulge}
\label{bulgenum}

The M31 bulge is assumed to be spherical, isotropic and non-rotating. This
relatively crude model is adequate because the bulge contributes only a modest
background correction in the region $R \lesssim 2\arcsec$ that is dominated by
the nucleus. We use the photometric parameters of KB99 for our bulge model. At
radii $<300\arcsec$, they found that the surface brightness of the bulge is
well-described by a \citet{ser68} law, $I(R)=I_0 \exp [-(R/R_n)^{1/n}]$, with
$I_0=15.40\hbox{ mag arcsec}^{-2}$ in $V$, $R_n=14\,\farcs0$, and
$n=2.19$. With this bulge-nucleus decomposition, the central surface
brightness of the bulge is 16\% of the peak surface brightness of the nucleus.

Much of our analysis will be based on the bulge-subtracted kinematic data of
KB99, in which case the bulge kinematics are irrelevant for our
models. However, for some purposes we compare the model to kinematic data that
includes the bulge light, and in this case we must model the bulge kinematics.
To do so, we assume that the bulge has constant mass-to-light ratio
$\Upsilon$, compute the gravitational potential of the spherical S\'ersic
profile described above, and add a central point mass corresponding to the
(unknown) total mass of the BH plus nuclear disk. We then solve for the
line-of-sight velocity distribution (LOSVD) at each radius, using formulae
from \citet{bt87} and \citet{sp98}.

We assume that the mass-to-light ratios $\Upsilon$ of the bulge and nuclear
disk are the same; this assumption is unlikely to be completely accurate since
the colors of the nucleus and bulge are different, but the model predictions
are only weakly dependent on the mass-to-light ratio of the bulge, and the
mass-to-light ratio of the disk only affects the model through the
displacement between the BH and the bulge center, which is at the center of
mass of the BH and the stellar disk.

\section{Numerical methods} \label{nummeth}

\subsection{Constructing the nuclear disk} \label{disknum}

We simulate the nuclear disk using a Monte-Carlo method, typically with
$N=10^7$ particles (there is little point in larger simulations, since $N$ is
already comparable to the number of stars in the actual disk, and much larger
than the number of giant stars, which dominate the light). First we assign
semimajor axes $a$ to the particles so that the distribution is uniform in
$\int_0^a\Sigma(a')a'\,da'$. For a given semimajor axis, the
DF (\ref{df}) implies that the mass in a small interval of the other orbital
elements is 
\be 
dm \propto \exp\left[-\frac{[{\bf e}-{\bf e}_m(a)]^2}{2\sigma_e^2}\right]
\exp\left[-\frac{I^2}{2\sigma_I(a)^2}\right]\,d{\bf e}\sin I\,dI\,dM\,d\Omega.
\ee 
Thus, for each particle we choose $\Omega$ and $M$ from uniform probability
distributions over the range $[0,2\pi)$; $e_x$ and $e_y$ from Gaussian
distributions with means $e_m(a)$ and 0 respectively, and standard deviation
$\sigma_e$; and $I$ from a Gaussian distribution weighted by $\sin I$, with
standard deviation $\sigma_I(a)$. We obtain the eccentric anomaly $E$ by
numerical solution of Kepler's equation [the third of equations
(\ref{orb-x-coord})]. We then use equations (\ref{orb-x-coord}) and
(\ref{orb-x-vel}) to determine the position $(x',y',z')$ and velocity $(\dot
x',\dot y',\dot z')$, and equations (\ref{orb-disk}) and (\ref{disk-sky}) and
their time derivatives to determine the positions and velocities in the
sky-plane coordinate system.

\subsection{Photometry} \label{photmodel}

We fit the model to the deconvolved HST V-band (F555W) image presented in
L98. The intensity scale is calibrated using the zero-point given in L98. We
assume that the pixel containing the maximum luminosity in the blue cluster in
the P2 peak is the position of the BH, and this point is also the origin of
our coordinate systems. The excess brightness contributed by the UV cluster at
P2 in the data is clipped out prior to fitting, since we may assume that these
stars have a small mass-to-light ratio. We compare the data to the model on a
sky-plane grid with cell size equal to the pixel size,
$0\,\farcs0228=0.085\pc$, and a circular boundary of radius
$R_g=5\pc=1\,\farcs34$.

We add the surface-brightness contribution from the bulge by assuming
that the center of the bulge coincides with the center of mass of the
disk-BH system; thus the center of the bulge is not precisely at the
origin.

\subsection{Kinematics} \label{kinmodel}

We fit the kinematics to the bulge-subtracted spectroscopic data from KB99,
which combine high S/N with reasonably good spatial resolution. These data
have a scale of 0\,\farcs0864/pixel, a PSF with FWHM of 0\,\farcs64, and a
slit width $2s = 0\,\farcs35$. Two spectra were taken at position
angle $\hbox{PA}=50\arcdeg$ and two more at
$\hbox{PA}=55\arcdeg$. KB99 then shifted these four spectra to a
common center and co-added them for further analysis.

We reproduce this procedure in the model. We compute the LOSVD from
the nucleus on a sky-plane grid with cell size equal to the pixel
size, and then smooth the LOSVDs by convolving with the PSF given in
equation (3) of KB99 and a top-hat slit of width $2s$. We then add the
LOSVDs at $\hbox{PA}=50\arcdeg$ and $\hbox{PA}=55\arcdeg$. Following
KB99, we fit the LOSVDs with a fourth-order Gauss-Hermite expansion
\citep{vdmf93,ger93}: 
\be 
\label{losvd} {\mathrm{LOSVD}}(v) = \frac{\gamma}{\sqrt{2\pi\sigma^2}} 
e^{-(v-V)^2/2\sigma^2}\left[1+ h_3 H_3\left(\frac{v-V}{\sigma}\right) + h_4
H_4\left(\frac{v-V}{\sigma}\right)\right]. 
\ee 
The coefficients $h_3$ and $h_4$ parametrize the lowest order odd and even
deviations from Gaussian line profiles.  The velocity bins are $5\kms$ wide
and are weighted equally in the fit.  The kinematic parameters extracted from
the fit ($V,\sigma, h_3, h_4$) can then be compared with the observations from
KB99.\footnote{Note that the kinematic data in KB99 (their Table 2)
have as their origin the position where the rotation speed $V=0$ relative to
the systemic velocity of M31, which is shifted from our origin at the center
of P2B toward P1 by $0\,\farcs051\pm0.014$ according to KB99 or 0\farcs031
according to B01.} 

\subsection{Parameter fitting} \label{params}

We fit the sky-plane image to the model using the statistic 
\be
\chi_1^2 = \sum_{R\le R_g} w(i,j)\left[\mathrm{datagrid}(i,j) -
\mathcal{ A}\ \mathrm{modelgrid}(i,j)\right]^2,
\label{eq:chione}
\ee 
where $\mathcal{A}$ specifies the overall normalization of the
model, $w(i,j)$ is the weight assigned to pixel $(i,j)$, and $R_g$ is the
radius of the region we are fitting.

One potential drawback of this approach is that most of the weight comes from
the outer parts of the image, while the most critical photometric test of the
model is its ability to fit the observations near the center, in particular
along the P1-P2 axis. We have addressed this problem in two ways: (i) we have
used a weight function $w(i,j)=\sqrt{\mathrm{datagrid}(i,j)}$, which gives
more weight to the high-surface brightness regions in P1 and P2 (the form of
this weighting factor was selected after experimenting with a variety of
possibilities). (ii) We have computed a second statistic, $\chi_2^2$, which is
similar to $\chi_1^2$ but restricted to pixels in a slit (of width one pixel
on the model photometric grid) that extends through P1 and P2 to a distance
$\pm R_g$ from the origin. This statistic focuses on the difficult task of
fitting the double structure of the nucleus. 

We specify the fit to the kinematic data by a statistic $\chi_3^2$, which
measures the mean-square differences of $V$ and $\sigma$ between the model and
the KB99 bulge-subtracted data at 46 points along the PA=50\arcdeg--55\arcdeg
axis (Table 3 of KB99). Each data point is weighted by the observational
uncertainty given by KB99. Note that KB99 present velocity and dispersion
measurements extracted with two techniques: Fourier Quotient (FQ, their Tables
2 and 3) and Fourier Correlation Quotient (FCQ, their Tables 4 and 5). The
Gauss-Hermite fitting procedure outlined above mimics the FCQ algorithm more
closely; however, the FQ data are given at significantly finer spatial
resolution. The velocity differences between the two algorithms are $\leq 2$
km s$^{-1}$; the close agreement between the two sets of velocity and
dispersion data is illustrated in Figures~\ref{fig:rot} and
\ref{fig:disp}. Thus we choose to fit to the KB99 FQ data.

We determine the best-fit model by minimizing the statistic
$\chi_1^2+\lambda_1\chi_2^2+\lambda_3\chi_3^2$, where the relative
weights $\lambda_1$ and $\lambda_2$ are chosen by hand such that for
the initial state model for the full photometry and kinematics fit,
$\chi_1^2 \sim \lambda_1\chi_2^2$ and $(\chi_1^2+\lambda_1\chi_2^2)
\sim \lambda_3\chi_3^2$. The procedure used to determine the initial
state model is described later in this section.

We do not fit to KB99's measurements of the Gauss-Hermite parameters $h_3$ and
$h_4$, or to spectroscopic data from other investigators. However, we shall
compare the best-fit models we obtain in this way to STIS observations of
kinematic parameters including $h_3$ and $h_4$. 

To reduce the dimensionality of our large parameter space, we choose two
parameters by hand: $c_1$ and $\Upsilon=M/L_V$. The Fermi cut-off parameter
$c_1$ (eq. \ref{surfden}) does not significantly affect the fit within the
inner 2\arcsec, and the photometry is quite insensitive to its exact value, so
we simply choose $c_1=4\pc^{-1}$. The mass-to-light ratio $\Upsilon$ does not
enter the kinematic fit since we use bulge-subtracted kinematic data, and only
affects the photometric fit through the displacement between the BH and the
bulge center, which is the barycenter of the BH and nuclear disk. To find the
position of the barycenter, we simply assume the same mass-to-light ratio for
the nuclear stars as the KB99 value for the bulge stars, $\Upsilon_V=5.7$. We
then use the resulting disk mass and the model's BH mass to compute the
position of the barycenter.

If the nuclear disk is assumed to be aligned with the M31 disk (``aligned
models'') we now have 11 free parameters: one orientation parameter
($\theta_a$, the angle in the disk plane between the sky plane and the
symmetry axis of the nuclear disk); one mass parameter (the BH mass
$M_\bullet$), and 9 parameters for the disk DF ($a_0, c_2, \sigma_I^0, a_I,
\alpha, a_g, a_e, w, \sigma_e$). If the nuclear disk orientation is fitted to
the data (``non-aligned models''), we must fit two more orientation
parameters, $\theta_i$ and $\theta_l$. 

The fitting is done with a downhill simplex algorithm \citep{pre92}.

To obtain initial conditions for the fitting program, we made use of the thin
disk model. It is relatively easy to find parameters for equations
(\ref{eccform}) and (\ref{surfden}) that give a thin disk $x$-axis
surface-brightness profile (\ref{thindisk}) similar to the observed profile
along the P1-P2 axis. Hence, reasoning that thickening the disk would lower
the surface brightness at P1, we chose as a starting point an ``extreme''
thin disk model which maximized the ratio of surface brightnesses P1/P2.

This extreme thin disk model produced optimum values for the parameters [$a_0,
a_1, \alpha, a_g, a_e, w$]. These were then used as a first guess to fit to
the photometry of the three-dimensional disk model (using only the statistic
$\chi_1$), with plausible initial guesses for $a_I$, $\theta_a$, $\sigma_I^0$
and $\sigma_e$. When the fitting algorithm converged for this limited
parameter set, they were then input into the full kinematic and photometric
fit as an initial guess, adding $M_\bullet$ to the fit.

It should be emphasized that the thin disk model was only used to
obtain the initial conditions; the rest of the fitting process uses
the thick disk model of \S\ref{eccdiskmodel}.

\section{Results} \label{results}

\begin{figure}
\plottwo{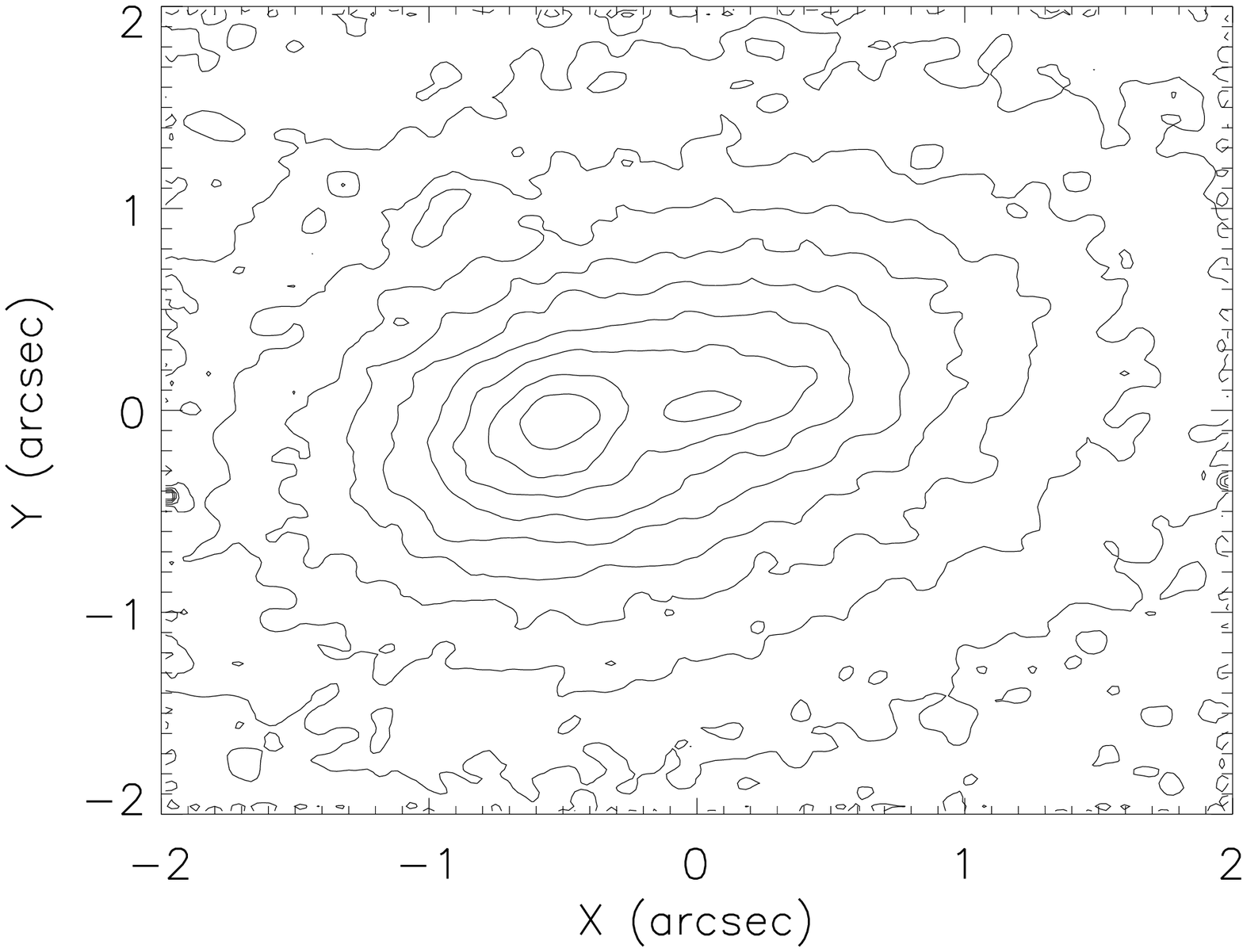}{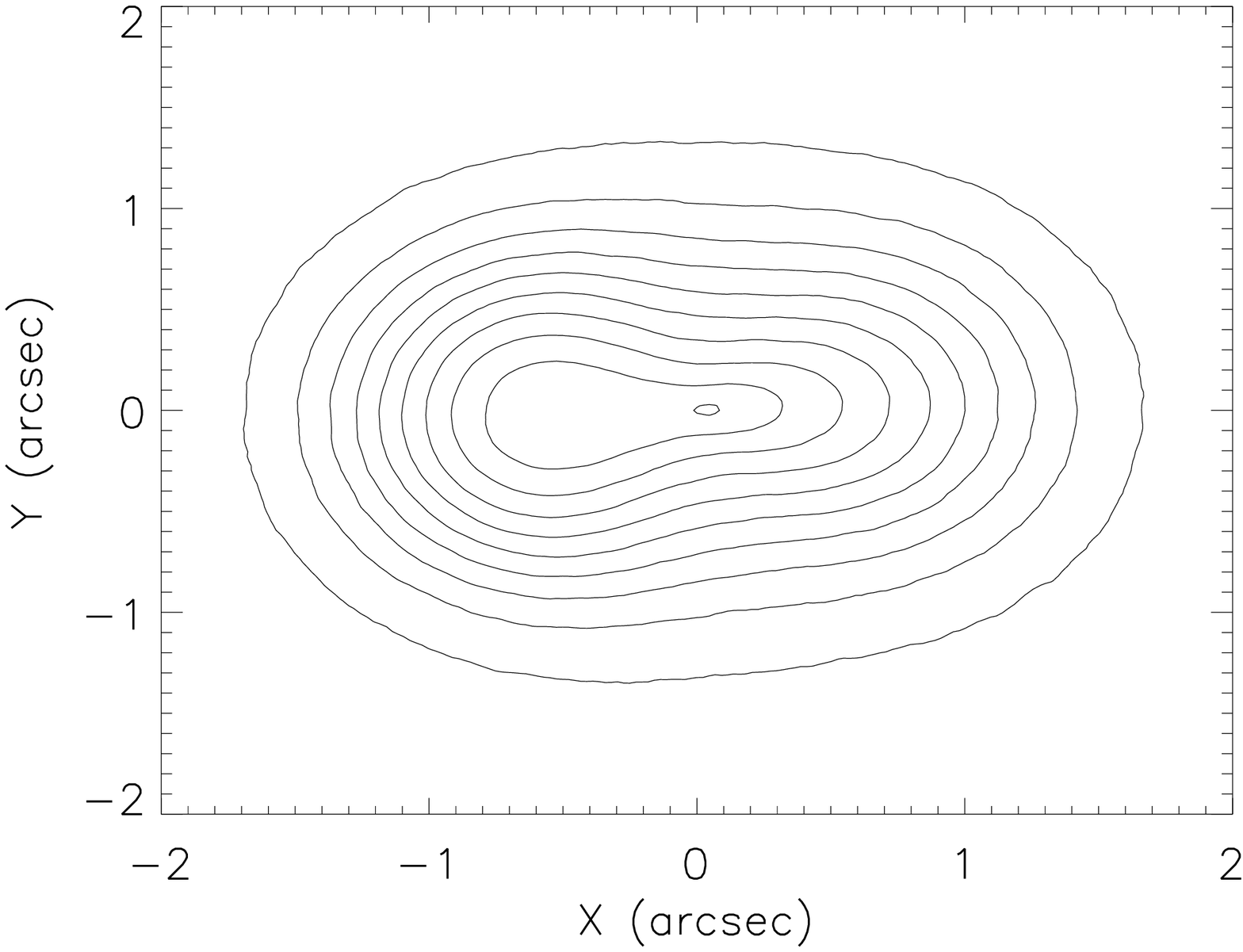}
\caption{Surface-brightness distribution in the nucleus of M31. The left panel
shows the data (V-band; from L98) and the right panel shows the best aligned
model. Contours are at 0.25 mag intervals. North is $55.7\arcdeg$ to
the left of the top of the plot.}
\label{fig:phota}
\end{figure} 

\begin{figure}
\plottwo{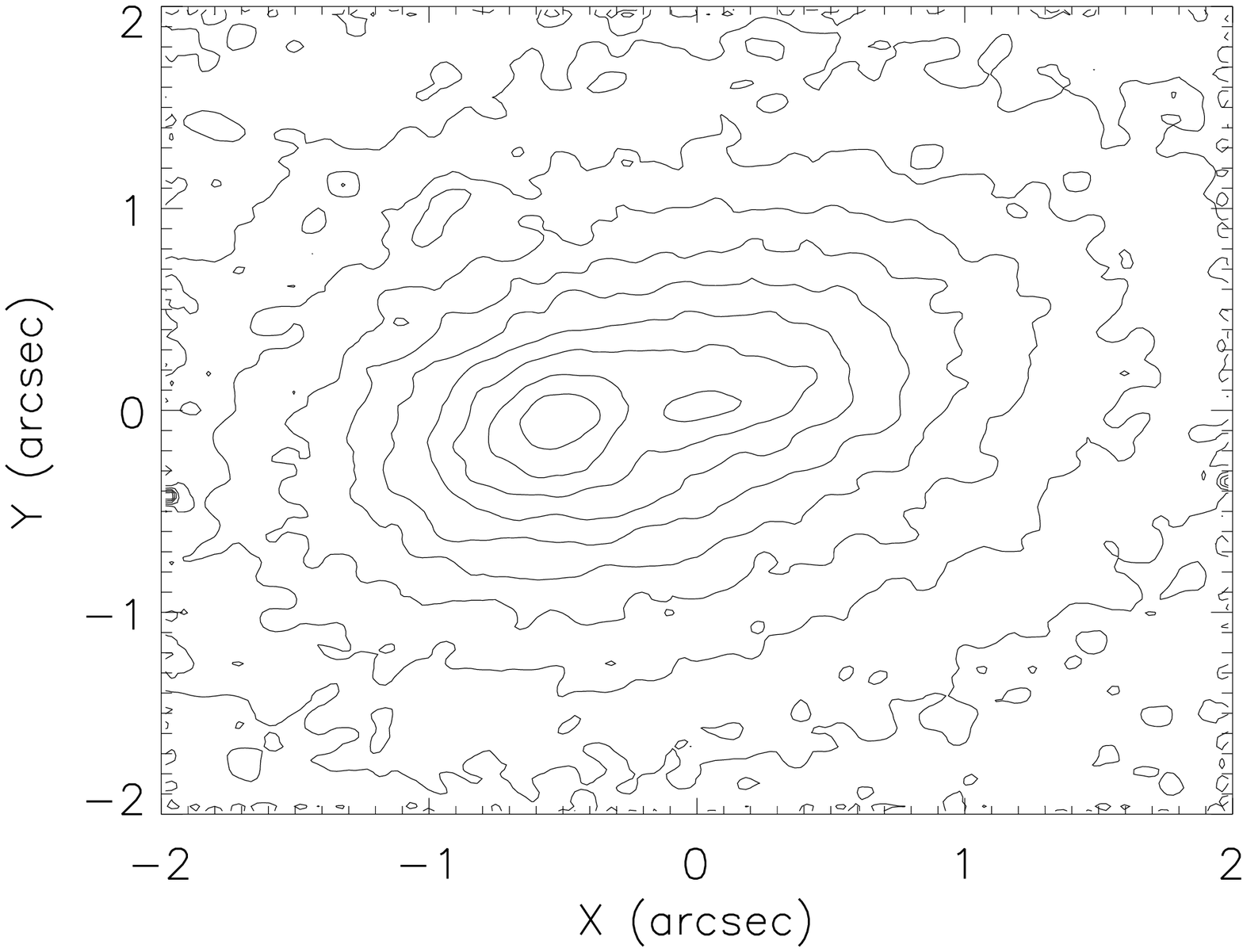}{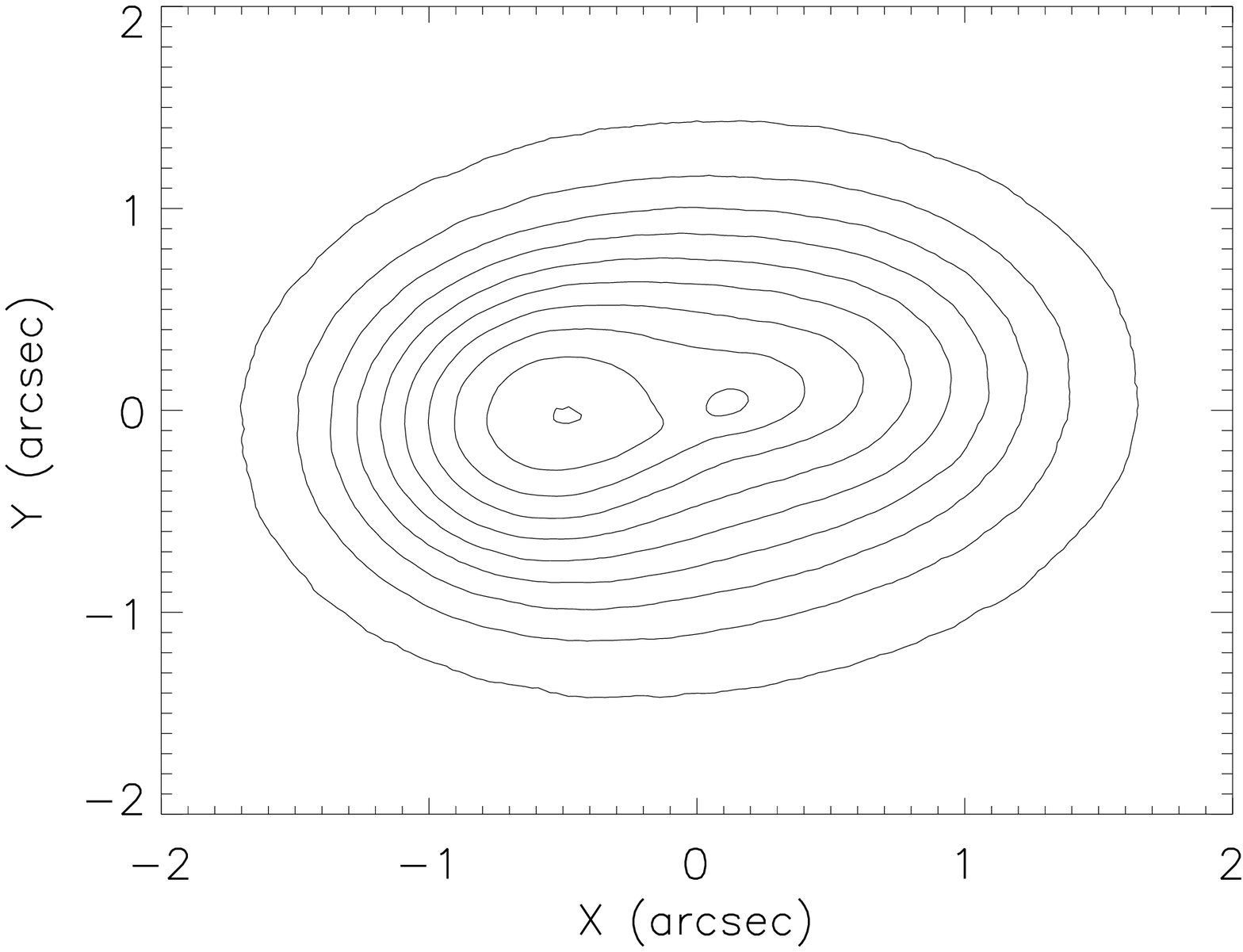}
\caption{Surface-brightness distribution in the nucleus of M31. The left panel
shows the data (V-band; from L98) and the right panel shows the best
non-aligned model. Contours are at 0.25 mag intervals. North is $55.7\arcdeg$
to the left of the top of the plot.}
\label{fig:photna}
\end{figure} 

\begin{figure}
\plottwo{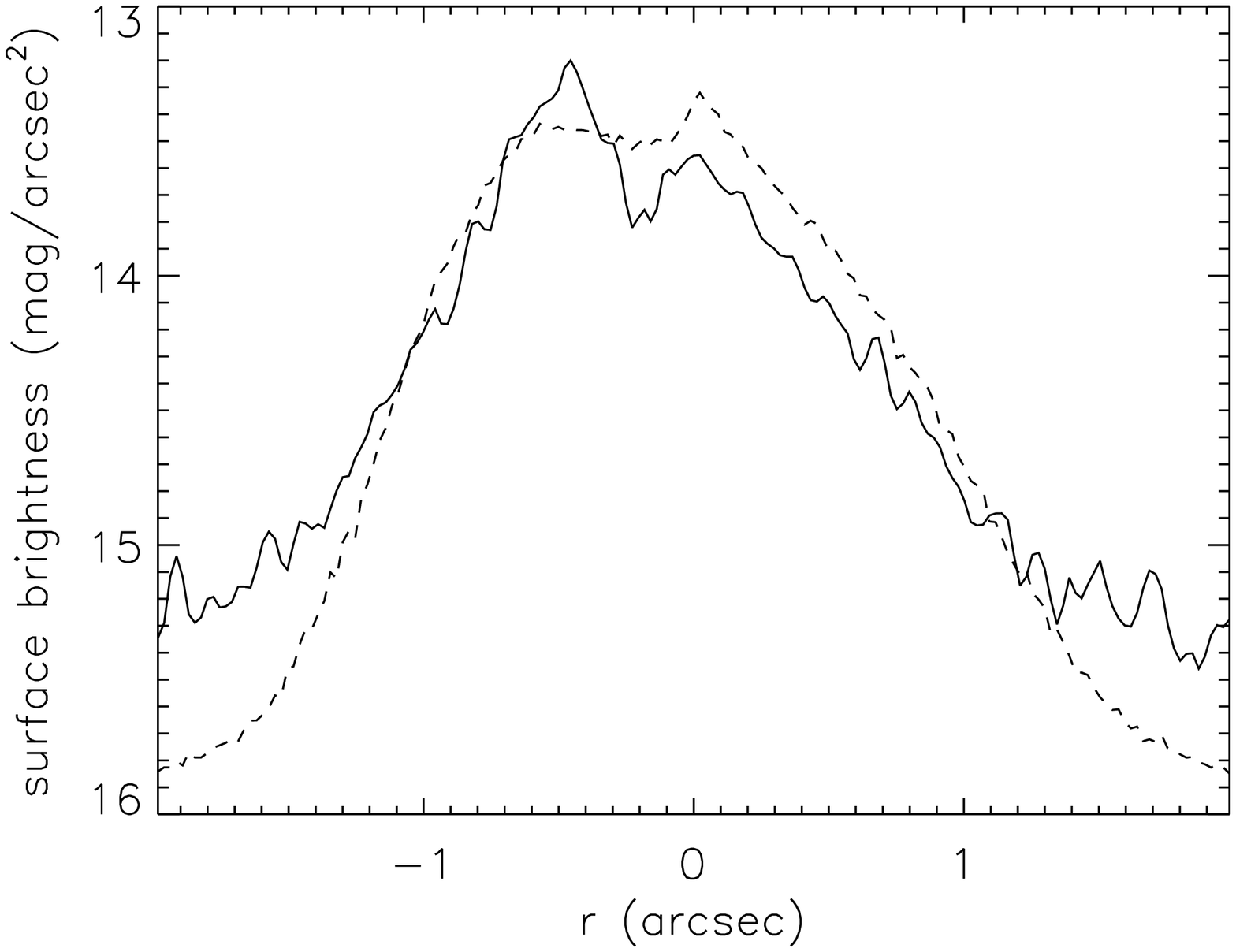}{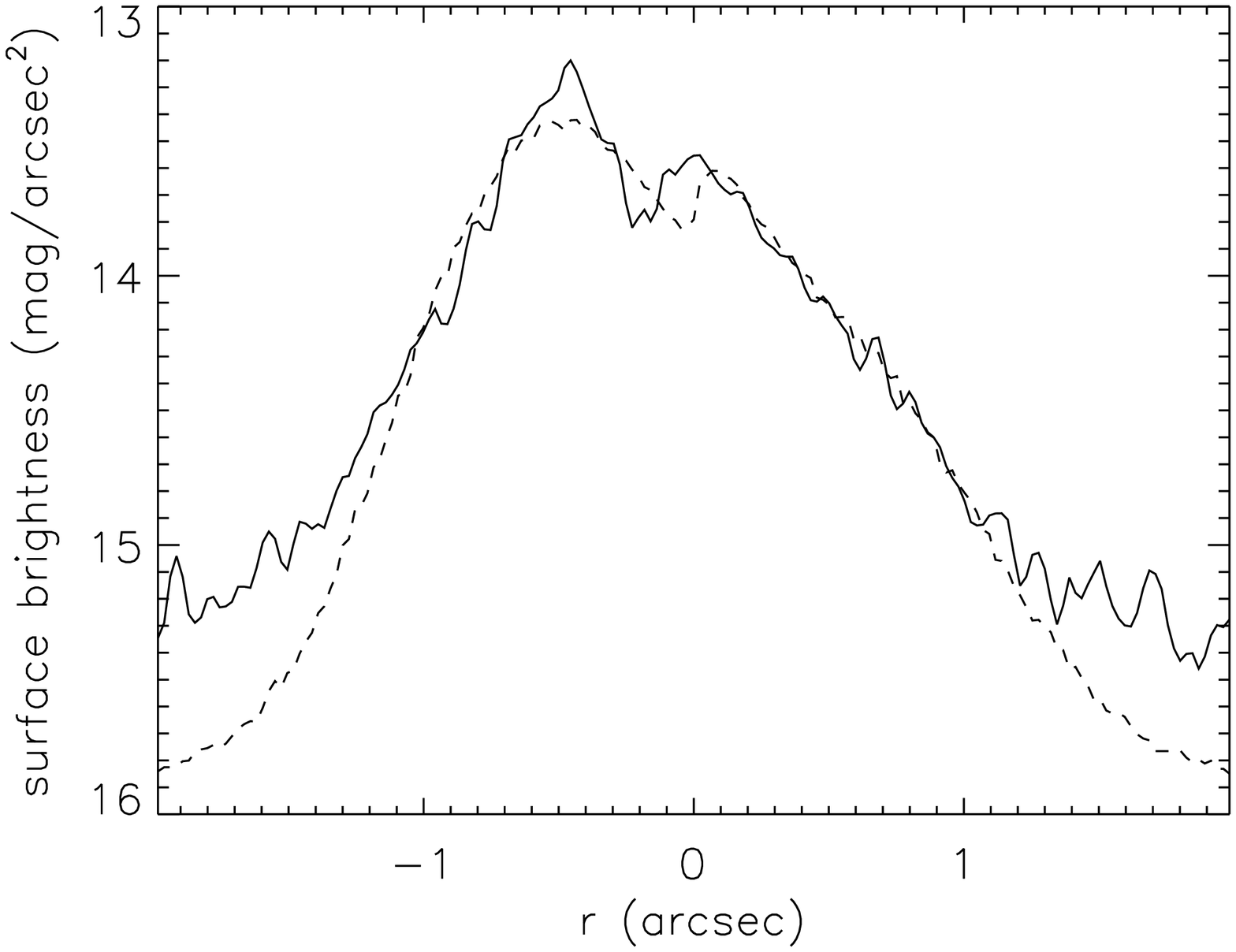}
\caption{Surface brightness along a slit of width 1 pixel ($=0\,\farcs0228$)
that extends through P1 and P2. The solid line represents the data (from L98)
and the dashed lines are the best aligned model (left panel) and non-aligned
model (right panel).}
\label{fig:slit}
\end{figure} 

\begin{figure}
\plottwo{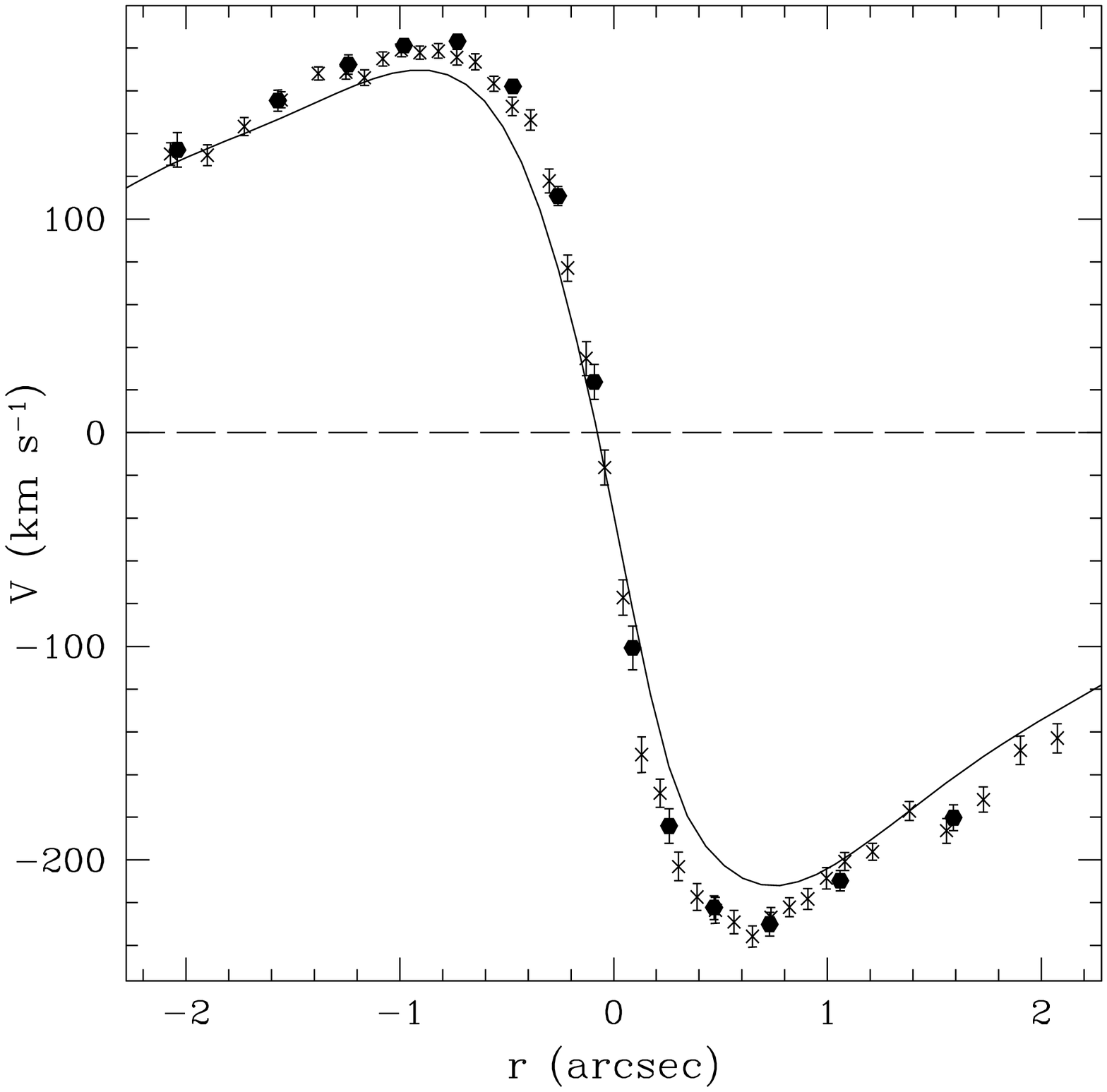}{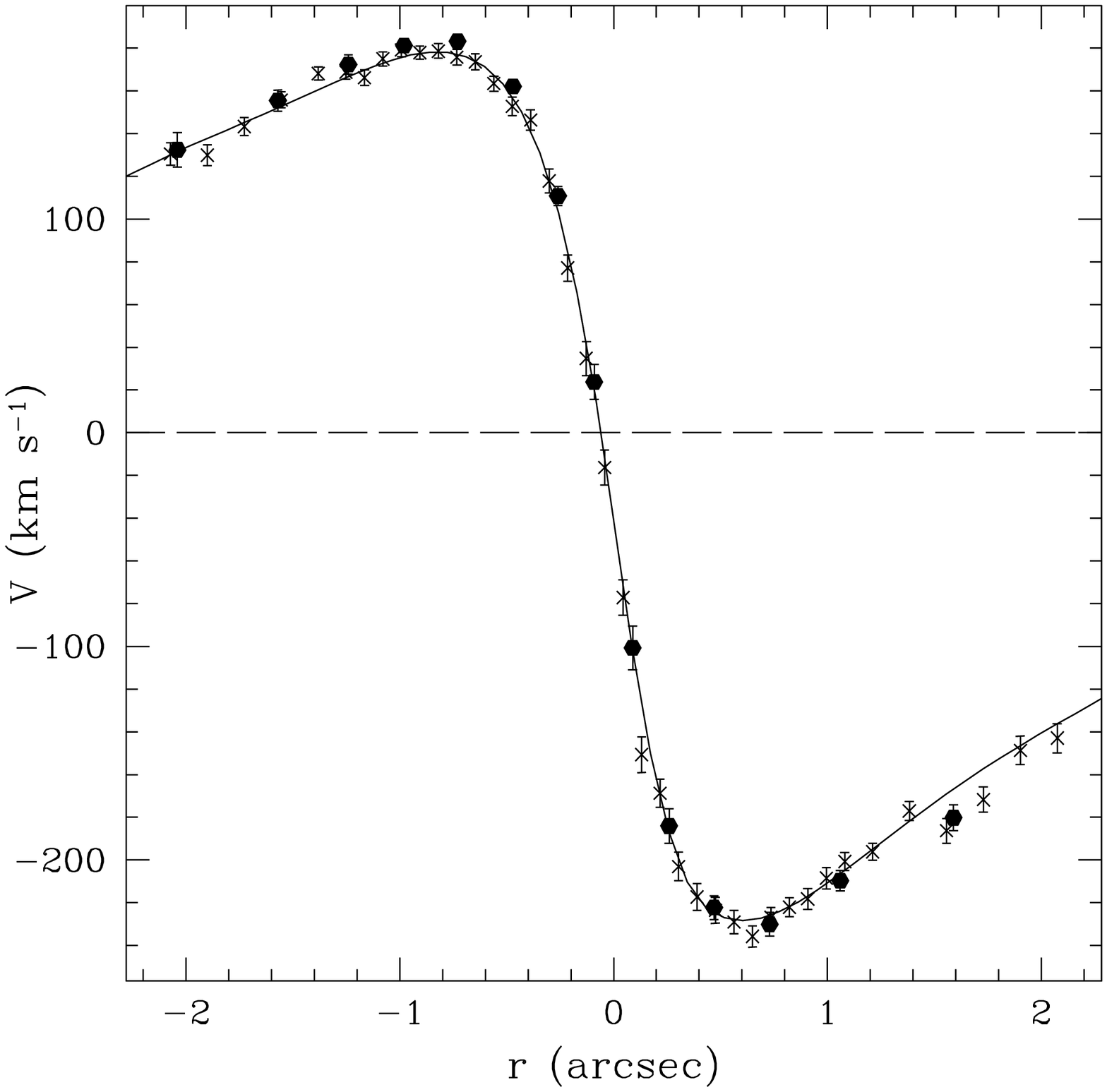}
\caption{Rotation speed $V$ along the axis at
PA=50$\arcdeg$--55$\arcdeg$, as determined by fitting the LOSVD to the
Gauss-Hermite expansion (\ref{losvd}). The data from KB99 after bulge
subtraction are shown as black dots (FCQ) and crosses (FQ), and the
solid lines are the best aligned model (left panel) and non-aligned
model (right panel).}
\label{fig:rot}
\end{figure} 

\begin{figure}
\plottwo{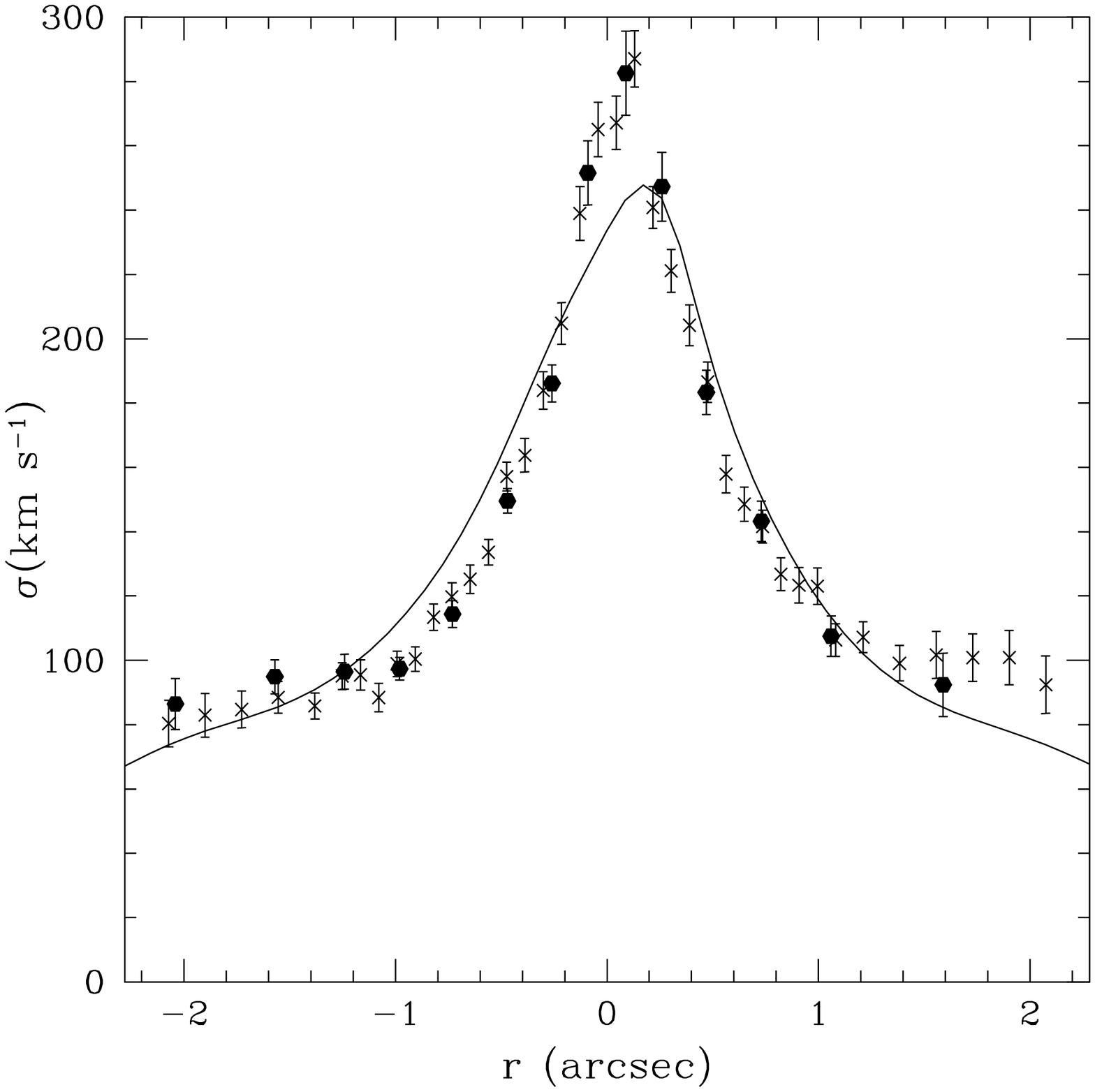}{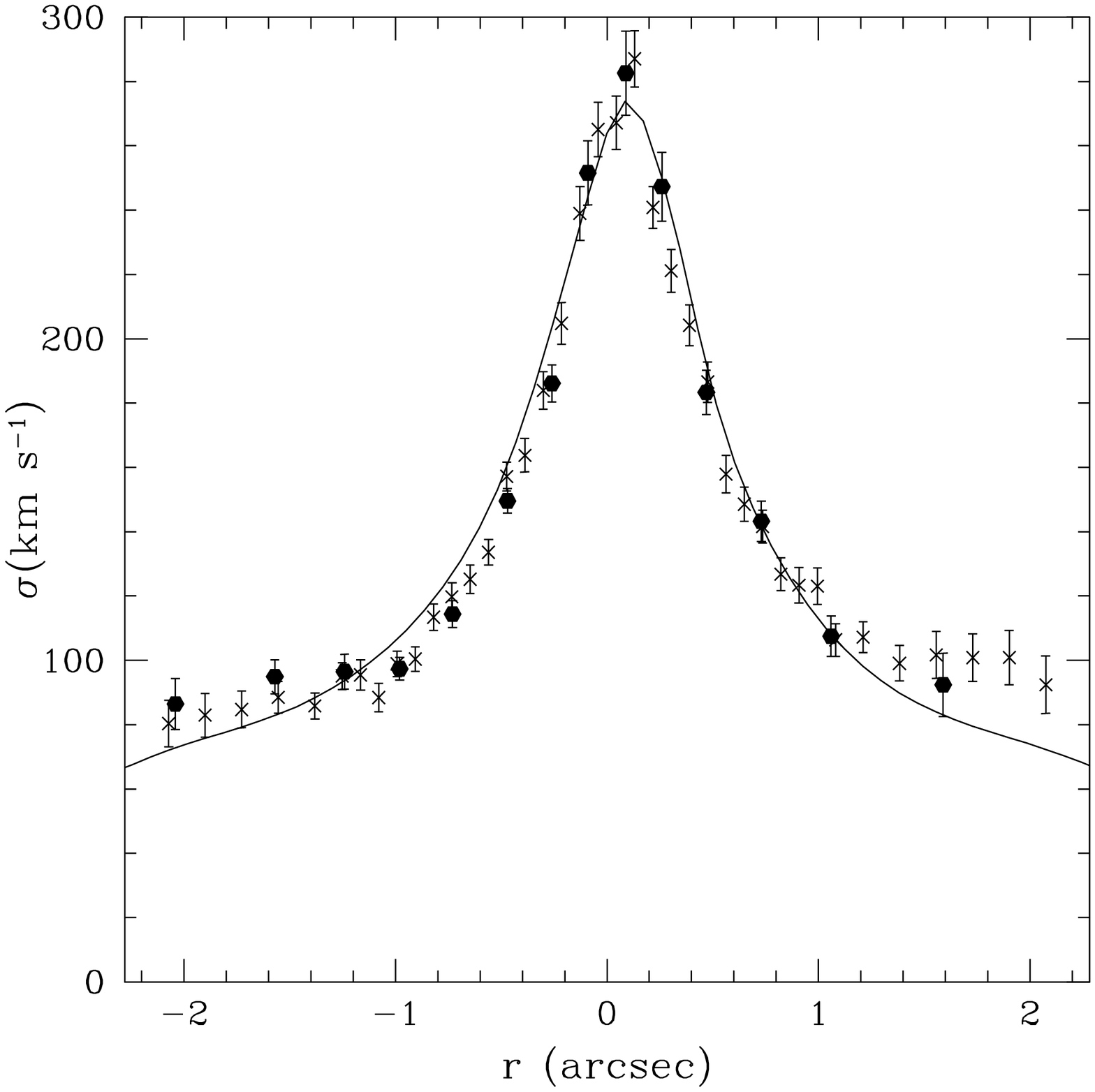}
\caption{The dispersion $\sigma$ along the axis at
PA=50$\arcdeg$--55$\arcdeg$, as determined by fitting the LOSVD to the
Gauss-Hermite expansion (\ref{losvd}). The data from KB99 after bulge
subtraction are shown as black dots (FCQ) and crosses (FQ), and the
solid lines are the best aligned model (left panel) and non-aligned
model (right panel).}
\label{fig:disp}
\end{figure} 

\begin{figure}
\plottwo{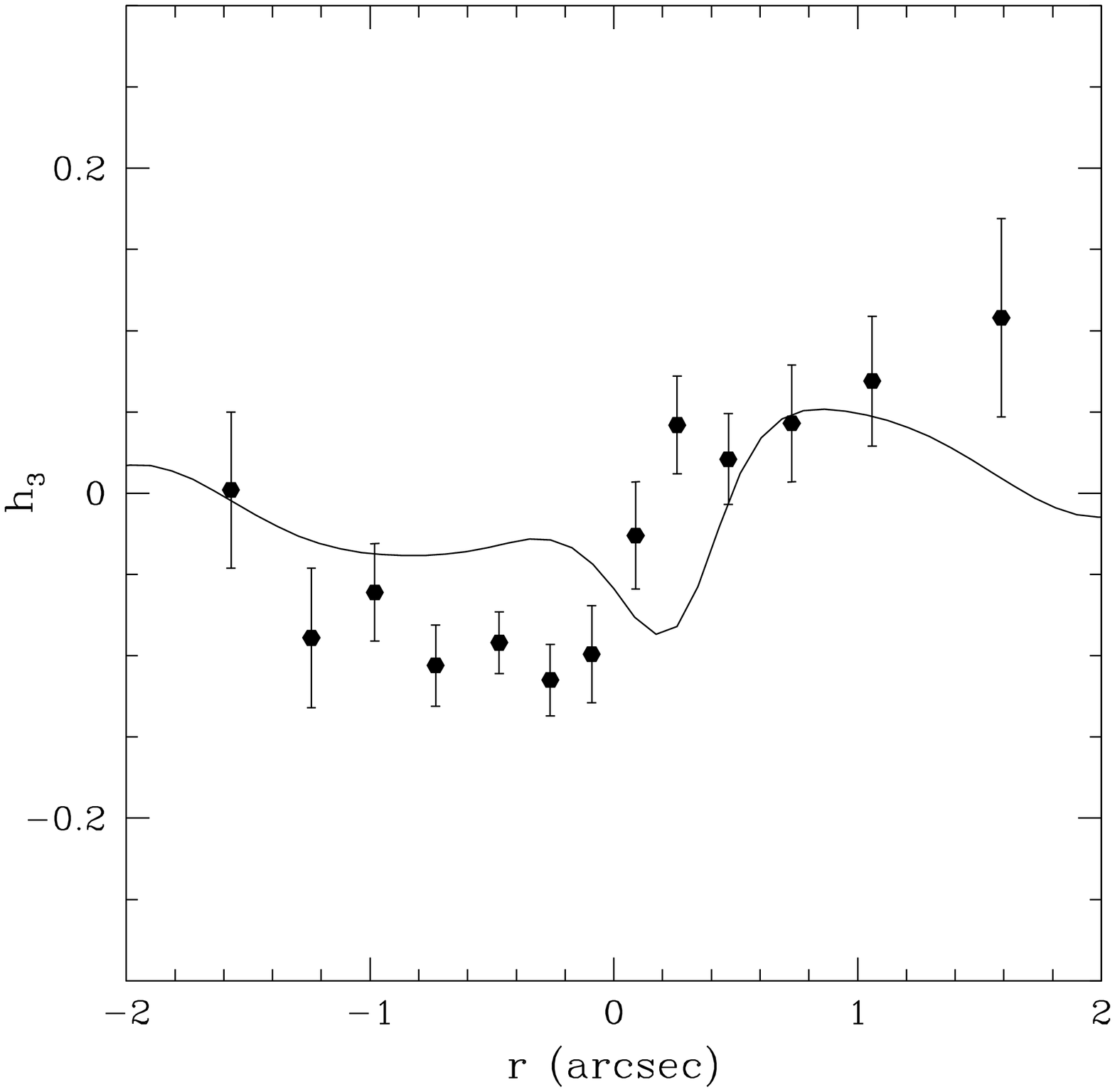}{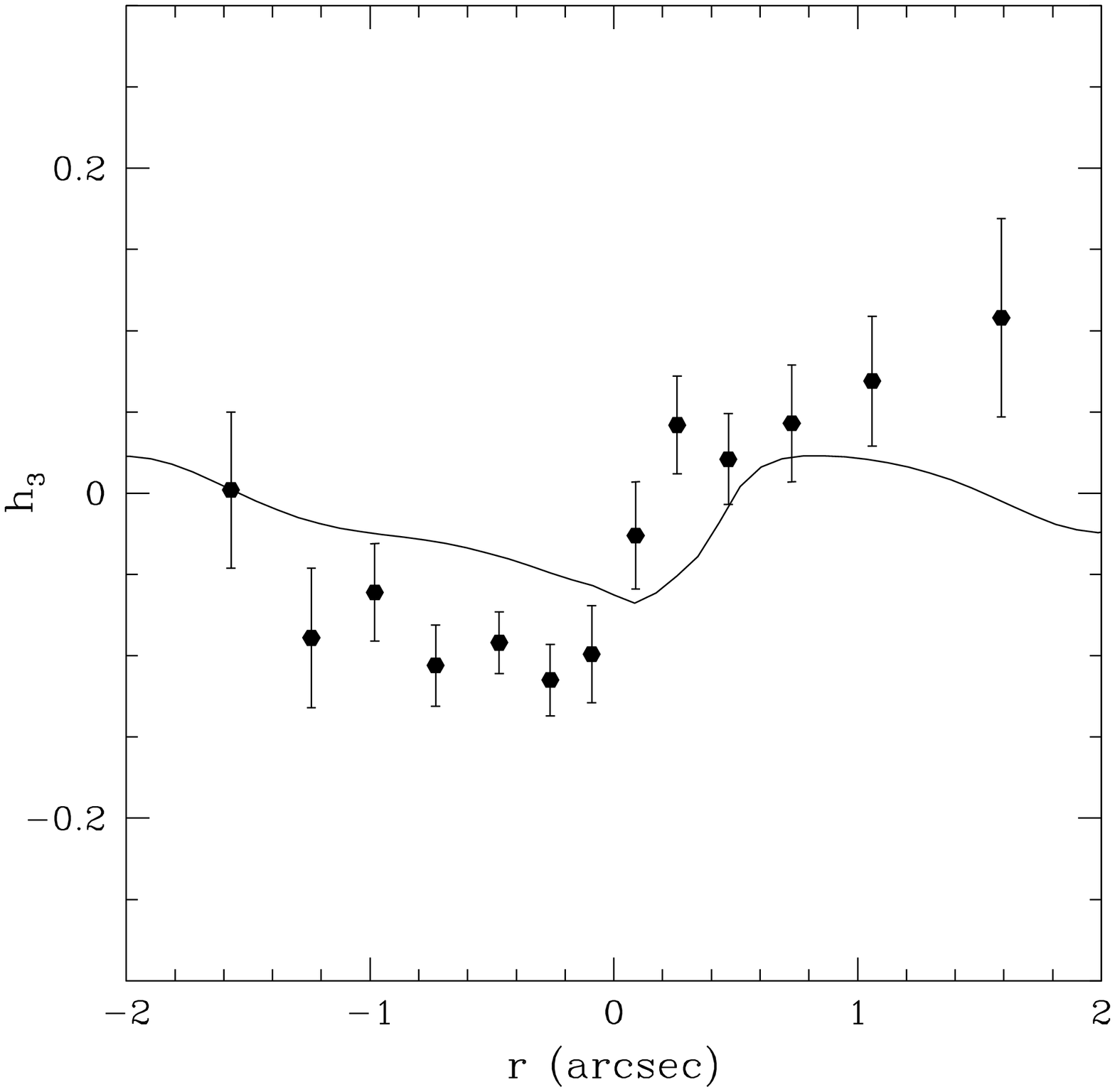}
\caption{The Gauss-Hermite parameter $h_3$ along the axis at
PA=50$\arcdeg$--55$\arcdeg$. The data from KB99 after bulge subtraction
are shown as black dots (FCQ), and the solid lines are the best
aligned model (left panel) and non-aligned model (right panel).}
\label{fig:hthree}
\end{figure} 

\begin{figure}
\plottwo{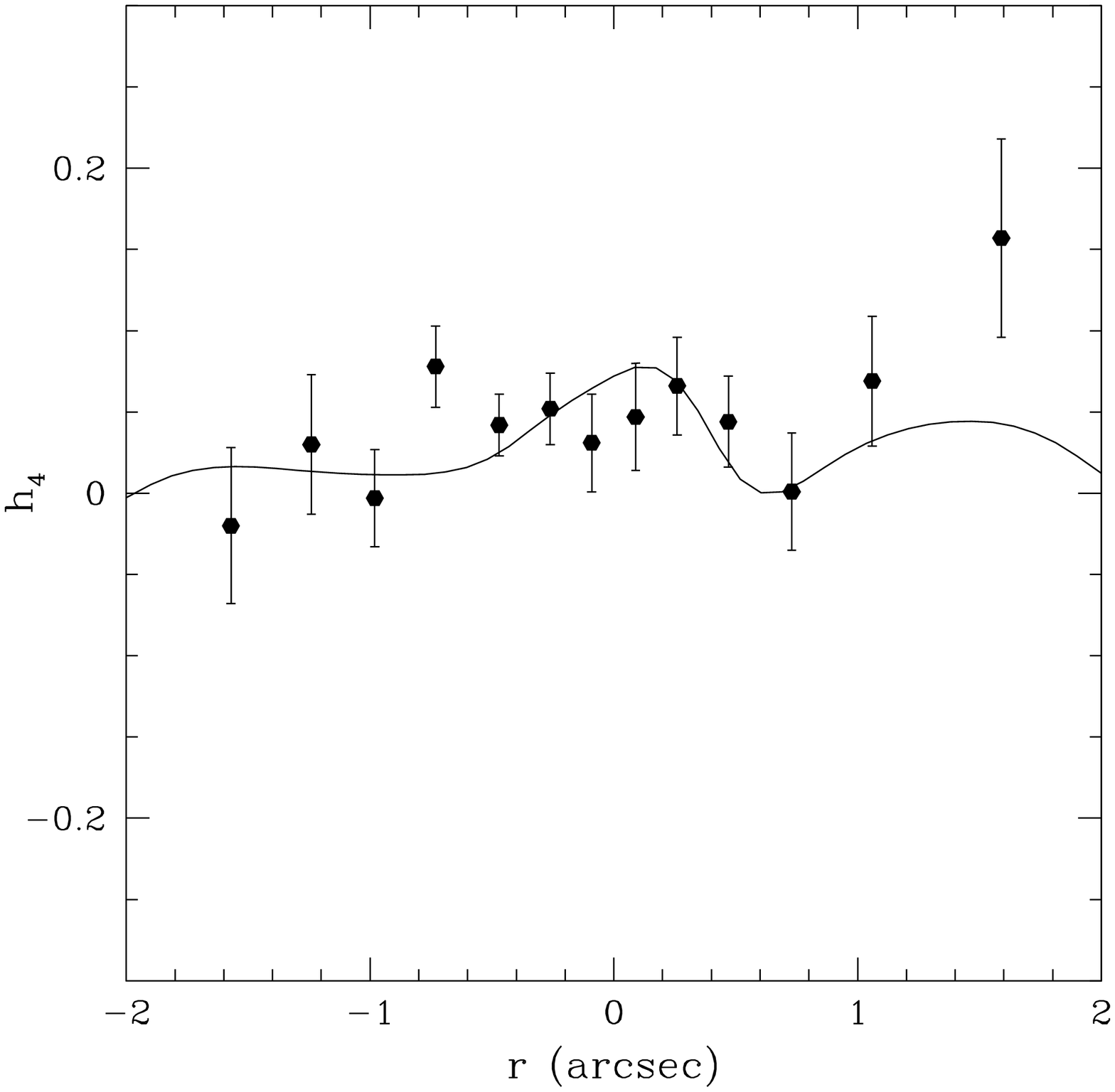}{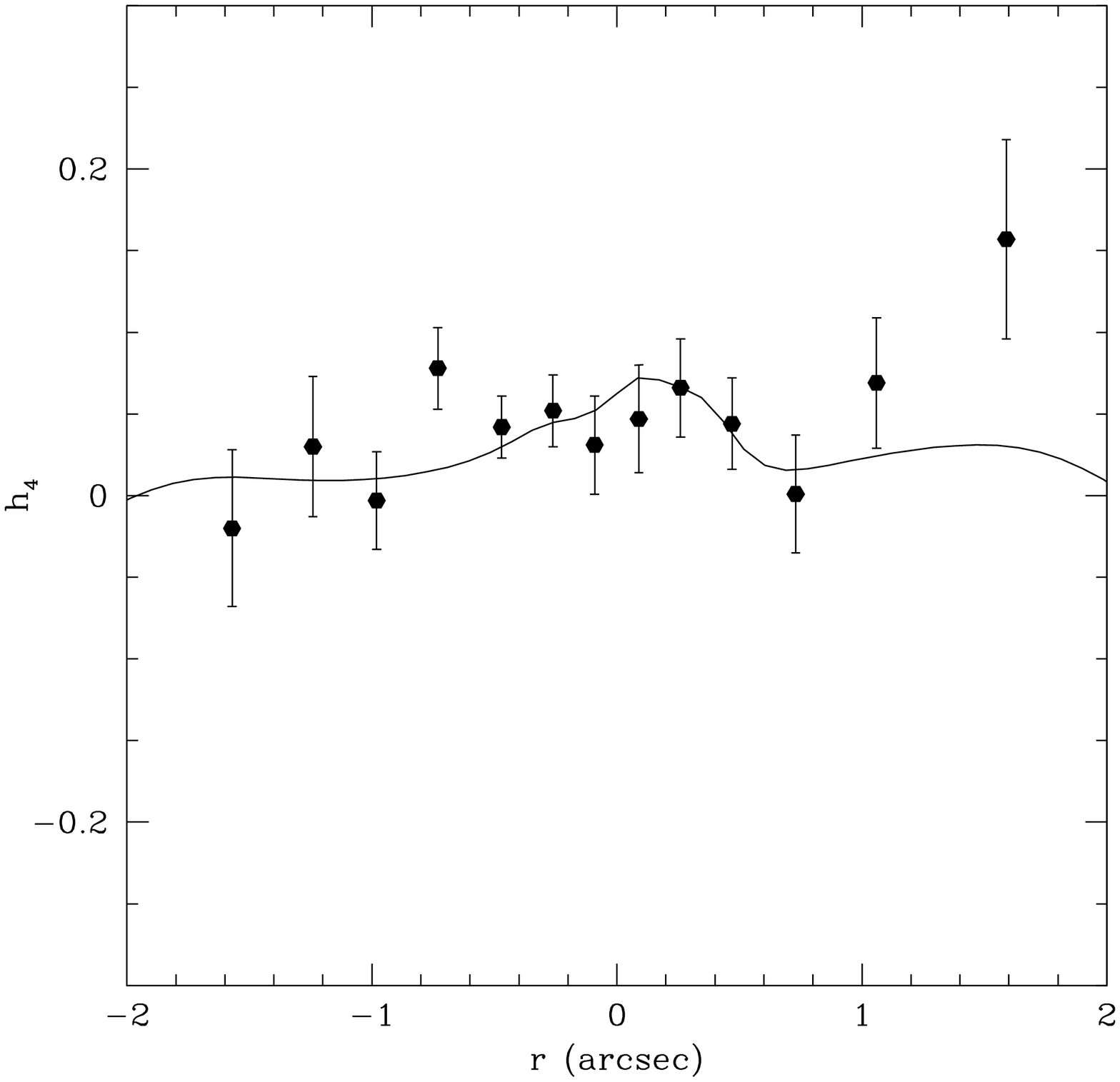}
\caption{The Gauss-Hermite parameter $h_4$ along the axis at
PA=50$\arcdeg$--55$\arcdeg$. The data from KB99 after bulge subtraction
are shown as black dots (FCQ), and the solid lines are the best
aligned model (left panel) and non-aligned model (right panel).}
\label{fig:hfour}
\end{figure}

\begin{deluxetable}{lcc}
\tablecolumns{3}
\tablewidth{0pt}
\tablecaption{Model parameters}
\tablehead{\colhead{parameter} & \colhead{aligned model} &
\colhead{non-aligned model}  \\}
\startdata
$\alpha$  & 0.288         &   0.197                \\
$a_e$ (pc)& 3.97          &   4.45                \\
$a_g$ (pc)& 1.51          &   1.71                \\
$w$ (pc)  & 1.53          &   1.52                \\
$M_\bullet (M_\odot)$ & $9.63\times10^7$ &   $1.02\times10^8$   \\
$a_0$ (pc)& 4.61          &   1.37                \\
$c_1$ (pc$^{-1}$) & 4\tablenotemark{a}     &   4\tablenotemark{a}        \\
$c_2$ (pc)& 3.79          &       4.24            \\
$\sigma_e$& 0.344         &       0.307            \\
$\theta_a$ (deg) & $-11.0$&       $-34.5$            \\
$a_I$ (pc)& 13.7          &       31.5            \\
$\sigma_I^0$ (deg) & 36.2 &      24.6            \\
$\theta_i$ (deg) & 77.5\tablenotemark{a}   &   54.1                \\
$\theta_l$ (deg) & $-52.3$\tablenotemark{a}&   $-42.8$                \\
$\Upsilon (M_\odot/L_\odot)$ &  5.7\tablenotemark{a} & 5.7\tablenotemark{a}  \\
\enddata
\tablenotetext{a}{This parameter was not fitted.}
\label{tab:parm}
\end{deluxetable}

\begin{deluxetable}{lccc}
\tablecolumns{6}
\tablewidth{0pt}
\tablecaption{Model photometry}
\tablehead{\colhead{parameter} &\colhead{A} &\colhead{N} & \colhead{data}\\}
\startdata
Separation of P1 and P2 (arcsec)  &   0.6     &     0.5    &  0.5 (L98) \\
P1--P2 position angle (deg)   &  $39\arcdeg$ & $42\arcdeg$ &  $43\arcdeg$ 
\\
Separation of P2 from bulge center (arcsec) & 0.04 &  0.04 & 0.07 (KB99) \\
Disk luminosity $\leq 1\arcsec$ projected 
($L_\odot$)  & $2.9 \times10^6$ & $2.9 
\times10^6$ & $2.9 \times10^6$ \\
Disk luminosity inside $4\arcsec \times 4\arcsec$ box 
($L_\odot$)  & $3.6 \times10^6$ & $3.6 
\times10^6$ & $5.4 \times10^6$ \\
\enddata
\tablecomments{A=aligned model; N=non-aligned model. Disk luminosity is
computed using bulge model from KB99. The projected model disk luminosity is
normalized to the projected bulge-subtracted data luminosity $\leq 1\arcsec$.} 
\label{tab:photres}
\end{deluxetable}

\begin{deluxetable}{lcccccc}
\tablecolumns{6}
\tablewidth{0pt}
\tablecaption{Model kinematics}
\tablehead{\colhead{parameter} & \colhead{A,S} &
\colhead{N,S} & \colhead{STIS (FCQ)} & \colhead{A,K} &
\colhead{N,K} & \colhead{KB99 (FQ)}   \\}
\startdata
P1 rotation maximum (km s$^{-1}$)      & 193    & 198    & 203 &  
170    &     178  &  179    \\
Position of rotation maximum (arcsec)  & $-0.76$  & $-0.61$  & $-0.59$  & 
$-0.95$   &   $-0.86$  &  $-0.99$  \\
Anti-P1 rotation minimum (km s$^{-1}$) & $-277$ & $-369$ & $-319$ & 
$-212$   &   $-228$ &  $-236$ \\
Position of rotation minimum (arcsec)  & 0.41  & 0.20  &  0.25 & 0.78  & 
0.60  &  0.65  \\
$V=0$ position (arcsec)                & $-0.17$  & $-0.12$ & $-0.12$ 
&$-0.05$   &   $-0.05$  &  $-0.05$  \\  
Dispersion peak (km s$^{-1}$)          & 252    & 328   &  302  & 
248    &     274  &  287   \\
Position of dispersion peak (arcsec)& 0.15    & 0.10    & 0.08 & 
0.17     &   0.09     &  0.13   \\
\enddata
\tablecomments{A=aligned model; N=non-aligned model; K=KB99 resolution;
S=STIS resolution}
\label{tab:kinres}
\end{deluxetable}

We shall describe two models: the best-fit aligned model, in which the nuclear
disk is assumed to lie in the symmetry plane of the M31 disk, and the best-fit
non-aligned model, in which the orientation of the nuclear disk is chosen to
optimize the fit.

The aligned model we describe here was obtained by minimizing the statistic
$\chi_1^2/(6\times10^6)+\chi_2^2/(6\times10^4)+\chi_3^2$, while the
non-aligned model was based on the statistic
$\chi_1^2/(1.2\times10^7)+\chi_2^2/(1.2\times10^5)+\chi_3^2$. The
parameters of the models are given in Table \ref{tab:parm}.  

Figures \ref{fig:phota} and \ref{fig:photna} show the surface-brightness
distributions in the data and the two models. Both models correctly
reproduce the apparent double structure of the nucleus, and the approximate
sizes and shapes of P1 and P2. However, in the aligned model the maximum
surface brightness of P1 is lower than the maximum brightness in P2, while in
the observations P1 is brighter. The non-aligned model does much better in
reproducing the relative surface brightnesses of P1 and P2. 

The position angle of the line joining P1 and P2 is $39\arcdeg$ in the aligned
model and $42\arcdeg$ in the non-aligned model, close to the observed value of
$43\arcdeg$. The position angle of the major axis of P1 in the data (as
determined by the isophotes with semimajor axis less than 0\,\farcs2) is
$63\arcdeg$. This $20\arcdeg$ isophote twist is not reproduced in the
models. However, the observed twist is strongly affected by the limited number
of giant stars that contribute to the surface brightness in the middle of
P1. We have explored this effect by reducing the number of stars in our
Monte-Carlo realization by a factor $\sim 100$ to approximate the actual
number of giants in the nucleus (of order $10^5$). We find that in this case
the position angle of the P1 major axis varies significantly in different
realizations; therefore, the difference in position angle between the major
axis of P1 and the P1-P2 axis does not provide much useful
information. However, the range of position angles seen in realizations of the
non-aligned model is larger than in the aligned model, so that the non-aligned
model is more easily able to accommodate the $20\arcdeg$ isophote twist seen
in the data.

Figure \ref{fig:slit} shows the surface brightness along the P1-P2 axis in
both models. In the data, the surface-brightness peak at P1 is approximately
0.4 mag brighter than the peak at P2. In the aligned model, P1 is fainter than
P2 by 0.1 mag, while in the non-aligned model P1 is brighter by 0.2 mag; thus
the fit of the non-aligned model to the relative brightnesses of P1 and P2,
though not perfect, is much better than the aligned model. The brightness of
P1 can be increased in the aligned model by reducing the disk thickness
(smaller $\sigma_I^0$) but then P1 becomes too elongated.

The non-aligned model correctly reproduces the dip in surface brightness
between P1 and P2, but the minimum is displaced 0\,\farcs2 further toward P2
than it is in the data. A feature of both models is that the P2 peak is
displaced to the anti-P1 side of the our origin at the BH (i.e. the UV peak
P2B). This is a consequence of the luminosity depression imposed at the origin
($\Sigma(a)\propto a^2$, eq.\ \ref{surfden}). While in principle this
displacement could be an important test of the model, with implications for
the kinematics (specifically, the displacement of the dispersion peak to the
anti-P1 side), we have found that this displacement cannot be determined
reliably if the number of stars in the simulation is reduced to approximate
the number of giant stars in the actual system: in this case Monte-Carlo
simulations can be created which have no significant displacement of the P2
peak from P2B. 

Note that both models fail to reproduce the surface brightness outside $\sim
1\,\farcs3$, because the surface brightness in the model falls off more
sharply than in the data. The obvious fix is to increase the cutoff parameter
$c_2$ in equation (\ref{surfden}).  However, this change affects the fit to
the kinematics, decreasing the amplitude of the rotation minimum and lowering
the dispersion peak. The discrepancy in the surface brightness at large radii
could arise either because the real bulge is cuspier than our model, which we
adopted from KB99, or because the nucleus contains extended wings that are not
captured by our parametrization. 

Figure \ref{fig:rot} shows the rotation speed $V$---more precisely, the
parameter $V$ in the Gauss-Hermite expansion (\ref{losvd})---along the average
of the two slits at PA=50$\arcdeg$ and 55$\arcdeg$ measured in KB99. The
aligned model, shown in the left panel, reproduces the approximate shape of
the rotation curve, in particular the strong asymmetry (the peak of the
observed rotation curve is at $179\kms$ on the P1 side and $236\kms$ on the
anti-P1 side; see Table \ref{tab:kinres} for the peak velocities in the
models). The zero-point of the model rotation curve is displaced from P2B
toward P1 by $0\,\farcs05$, consistent with the displacement of
0\,\farcs05--0\,\farcs10 estimated by KB99. The most significant discrepancy
is that the maximum rotation speed on the anti-P1 side, near 0\,\farcs6, is
too small by about $24\kms$ or 10\%; the rotation speed near the peak on the
P1 side is also too low, but by a smaller amount.

The right panel shows the rotation curve for the non-aligned model, which fits
the rotation curve substantially better than the best aligned model.

Figure \ref{fig:disp} shows the dispersion profile along the same slits. The
dispersion profile in the aligned model, shown on the left, fits the data
fairly well, but (i) the maximum dispersion is too low by 14\% ($248\kms$
compared to $287\kms$); (ii) the model dispersion is too large by about
$10\kms$ on the P1 side between radii of 0\,\farcs4 and 1\,\farcs1. The
dispersion is also too high near 2\arcsec\ on the anti-P1 side, but this
discrepancy probably is due to the shortcomings of the bulge model discussed
above.

The model correctly reproduces the displacement of the dispersion peak from
P2B by about 0\,\farcs13 in the anti-P1 direction---the displacement in the
model is about 0\,\farcs17. We could increase the peak dispersion in the
model by increasing the BH mass, but then the excess model dispersion on the
P1 side would become even worse.

We have attempted to increase the maximum dispersion by adding a compact
spherical stellar system at P2 (in the form of a Plummer sphere); the hope was
that the high-velocity stars in a small, low-luminosity system of this kind
would increase the dispersion without degrading the fit to the
photometry. However, we found that increasing the dispersion profile in this
manner made the rotation curve too symmetrical, so the overall fit was not
improved.

The dispersion profile and rotation curve of the aligned model could be made
to fit the KB99 data if we assume that the width of the PSF is 10--15\%
smaller than the value used by KB99. However, it is unlikely that there is an
error of this magnitude in the estimated PSF.

The right panel of Figure \ref{fig:disp} shows the dispersion profile in the
non-aligned model. The model profile fits the data substantially better than
the best aligned model, though the maximum dispersion is still too low by
5\%. The dispersion peak is displaced by 0\,\farcs09 from P2B in the anti-P1
direction, in good agreement with the observed displacement of 0\,\farcs13.

Figures \ref{fig:hthree} and \ref{fig:hfour} show the parameters $h_3$ and
$h_4$ in the Gauss-Hermite expansion. Both the aligned and non-aligned models
are moderately successful at reproducing the data within the large
uncertainties; thus it appears that a model that is fit to the
overall scale and center of the velocity distribution (measured by $\sigma$
and $V$) also reproduces the gross features of its shape. Note that we do not
use data on either $h_3$ or $h_4$ in the fitting process, so the comparisons
in these figures are a measure of the predictive power of the eccentric-disk
model rather than the quality of the fit.

\begin{figure}
\plottwo{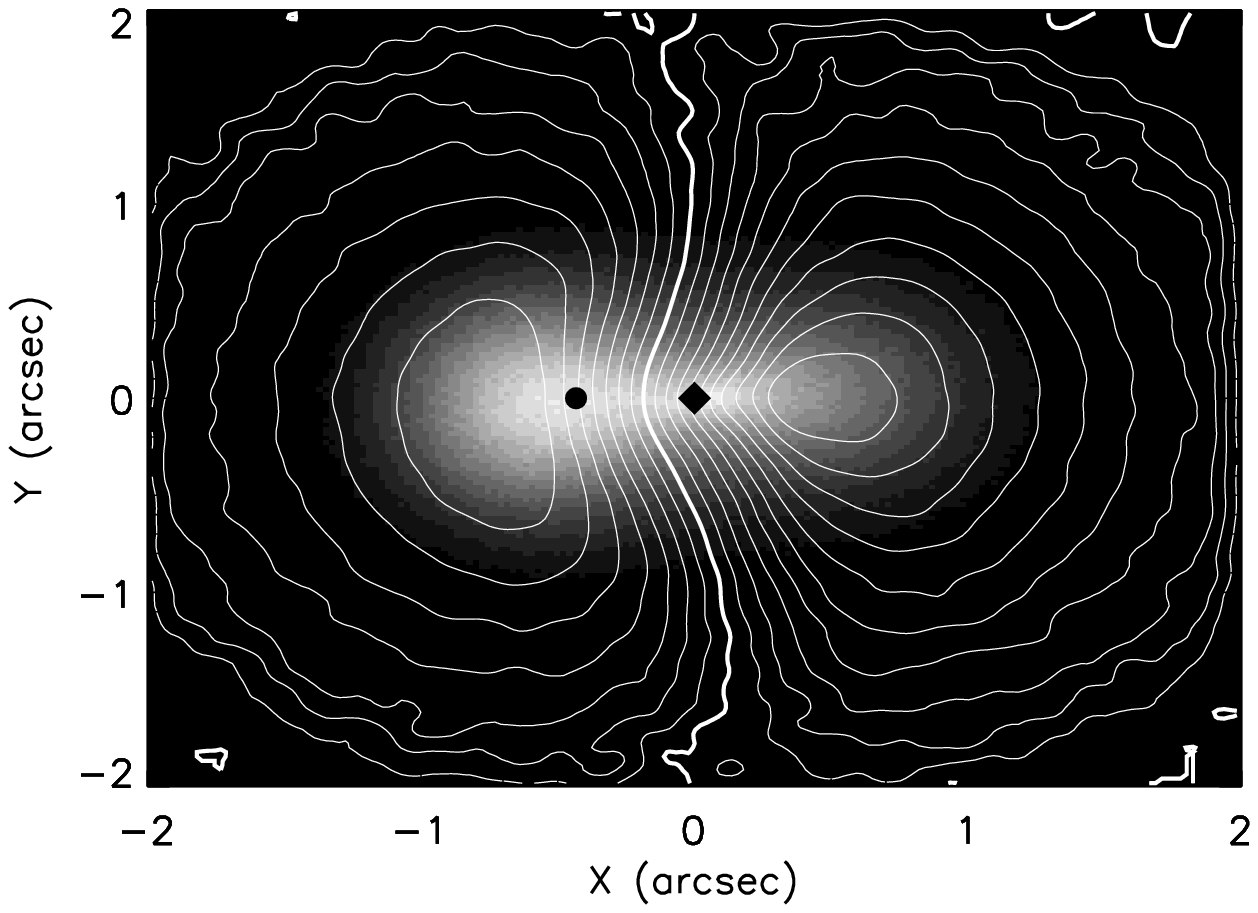}{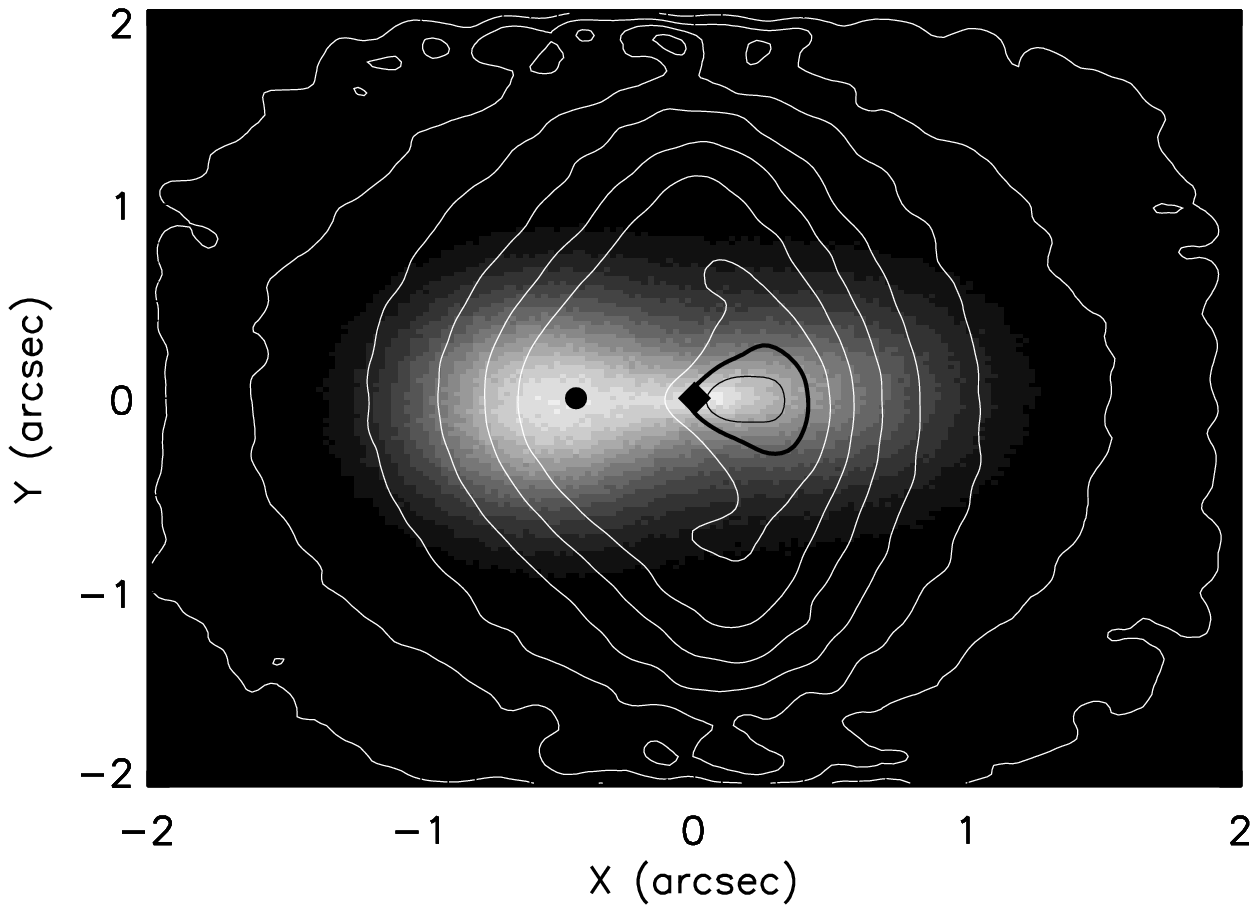}
\caption{Two-dimensional kinematics of the best aligned model. The left panel
shows the mean-velocity contours and the right panel shows the
velocity-dispersion contours; both panels are superimposed on a gray-scale
image of the nucleus. The contour interval is $25\kms$. The thick line is
the zero (systemic) velocity contour (left panel) and the $200\kms$ dispersion
contour (right panel). The locations of the center of P1 (filled circle) and
P2B (filled diamond at the origin) are shown. North is $55.7\arcdeg$ to the
left of the top of the plot.}
\label{fig:align2dkin}
\end{figure} 

\begin{figure}
\plottwo{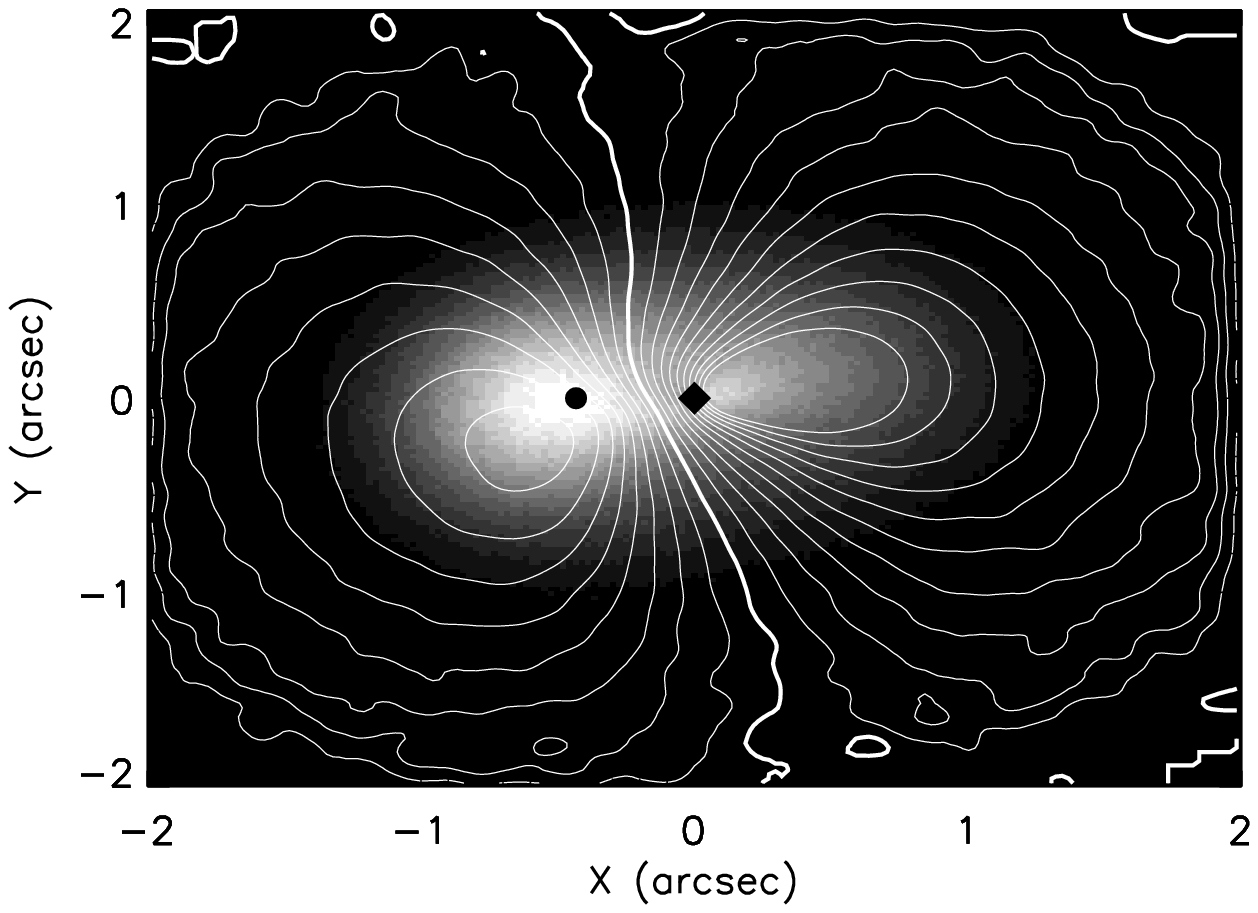}{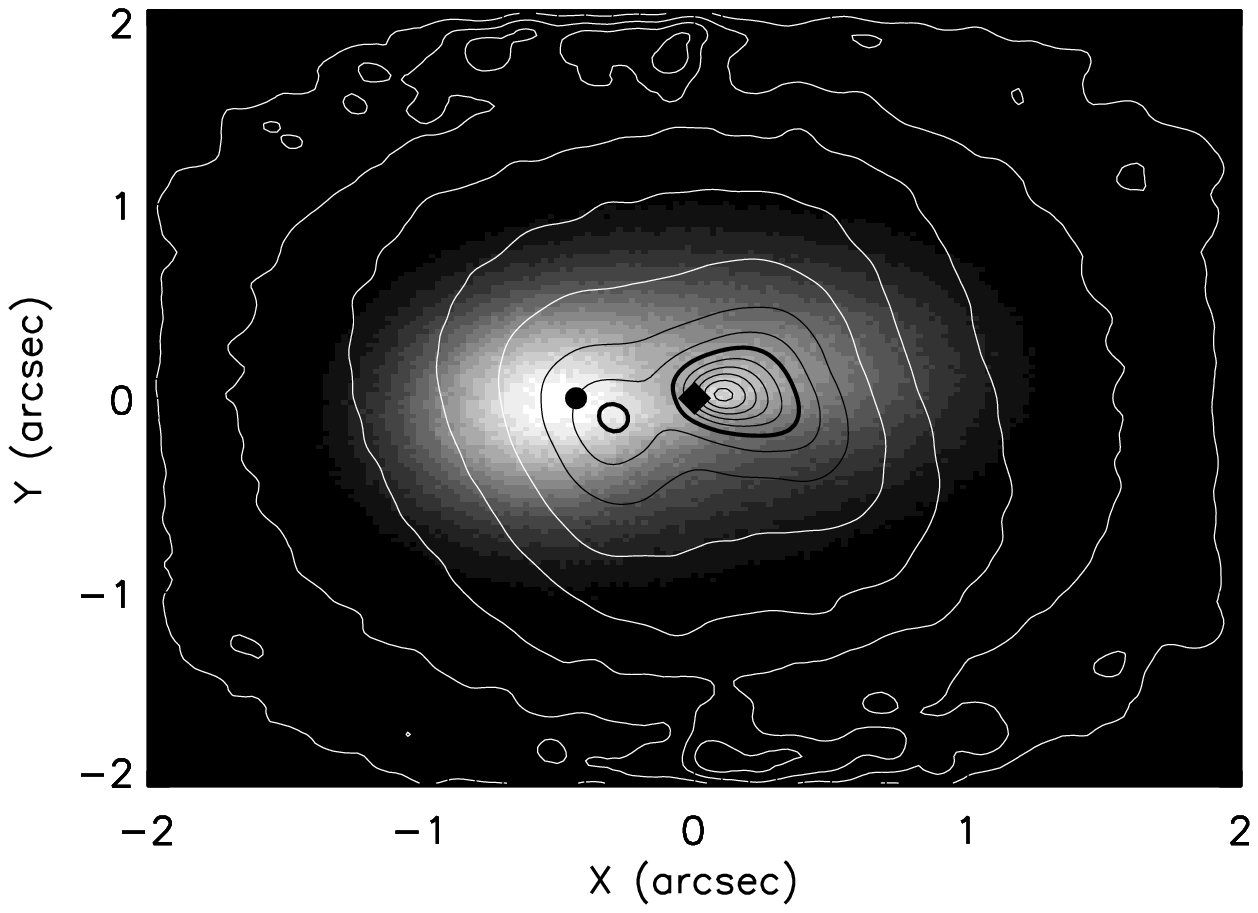}
\caption{Two-dimensional kinematics of the best non-aligned model. Parameters
as in Figure \ref{fig:align2dkin}.}
\label{fig:nonalign2dkin}
\end{figure} 

B01 have used the OASIS integral-field spectrograph, assisted by adaptive
optics, to measure the two-dimensional mean-velocity and velocity-dispersion
fields in the central 1\,\farcs5 of M31, with a PSF having FWHM of
$\sim$0\,\farcs4--0\,\farcs5. They find that the kinematic axis (the axis of
symmetry of the mean-velocity field) has a position angle of $56\arcdeg$, which
is rotated from the P1--P2 axis at position angle $43\arcdeg$. Figures
\ref{fig:align2dkin} and \ref{fig:nonalign2dkin} show the mean-velocity and
velocity-dispersion fields for the aligned and non-aligned models, as viewed
with a Gaussian PSF having FWHM of 0\,\farcs13 to smooth out the shot noise in
the contours. These figures can be directly compared to the model photometry
in Figures \ref{fig:phota} and \ref{fig:photna}, which has the same resolution
and orientation. Once again, the non-aligned model provides a better fit to
the data than the aligned model. In particular, the kinematic major axis
(defined by the line joining the velocity extrema) in the non-aligned model
has PA=48--50$\arcdeg$, much closer to the observed orientation of $56\arcdeg$
than the kinematic major axis in the aligned model, PA=40$\arcdeg$. We suspect
that the orientation of the kinematic axis is resolution-dependent, which may
account for the difference of $7\arcdeg$ between the non-aligned model and the
B01 data. In Figure 7 of B01, the isovelocity contours between P1 and P2 are
nearly perpendicular to the line joining the velocity extrema, while this
angle is oblique in our non-aligned model. We find that lower-resolution maps
of the velocity field of the non-aligned model yield isovelocity contours that
are more nearly perpendicular to the kinematic axis, consistent with the B01
data.

\subsection{Comparison with STIS data} \label{stiscomp}

We now compare the predictions of the models described above to new
high-resolution STIS observations \citep{ben03}. Recall that our models were
fitted to the ground-based spectroscopic data from KB99; here, the kinematics
have simply been recomputed for STIS resolution (slit width 0\,\farcs1)
without modifying the fit. The effects of the instrumental broadening on the
LOSVD have been accounted for using a template star observed through the same
slit (R. Bender, private communication). We use the PSF given by
\citet{bow01}, which can be modeled as the sum of two Gaussians (K. Gebhardt,
private communication):
\begin{equation}
G(r) = 1.01 \exp\left[-\frac{r^2}{2(0.0329)^2}\right] + 0.12
\exp\left[-\frac{(r-0.105)^2}{2(0.0384)^2}\right],
\label{eq:stispsf}
\end{equation}
where $r$ is the distance in arcseconds. Note that a Wiener filter has been
applied to the STIS data (R.\ Bender, private communication) to remove 
high frequencies, a step that we have not taken in analyzing the
model kinematics. 

The spatial resolution of the STIS observations is far better than that of the
KB99 observations (the slit width is a factor of 3.5 smaller, and the FWHM of
the PSF is smaller by a factor of 5.4). Thus the comparison of the STIS
kinematics with our models is a stringent test of their predictive power. 

\begin{figure}
\plottwo{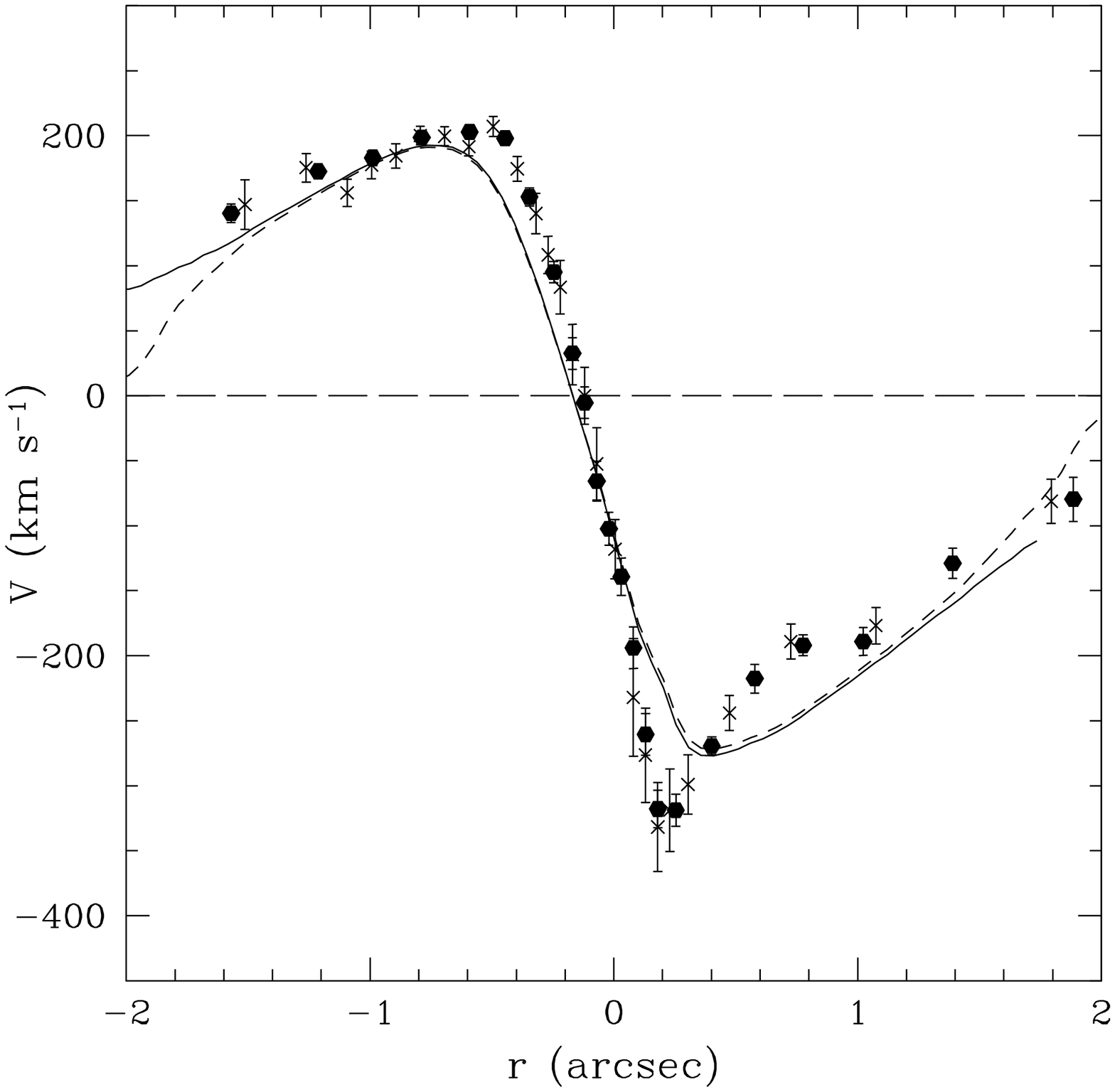}{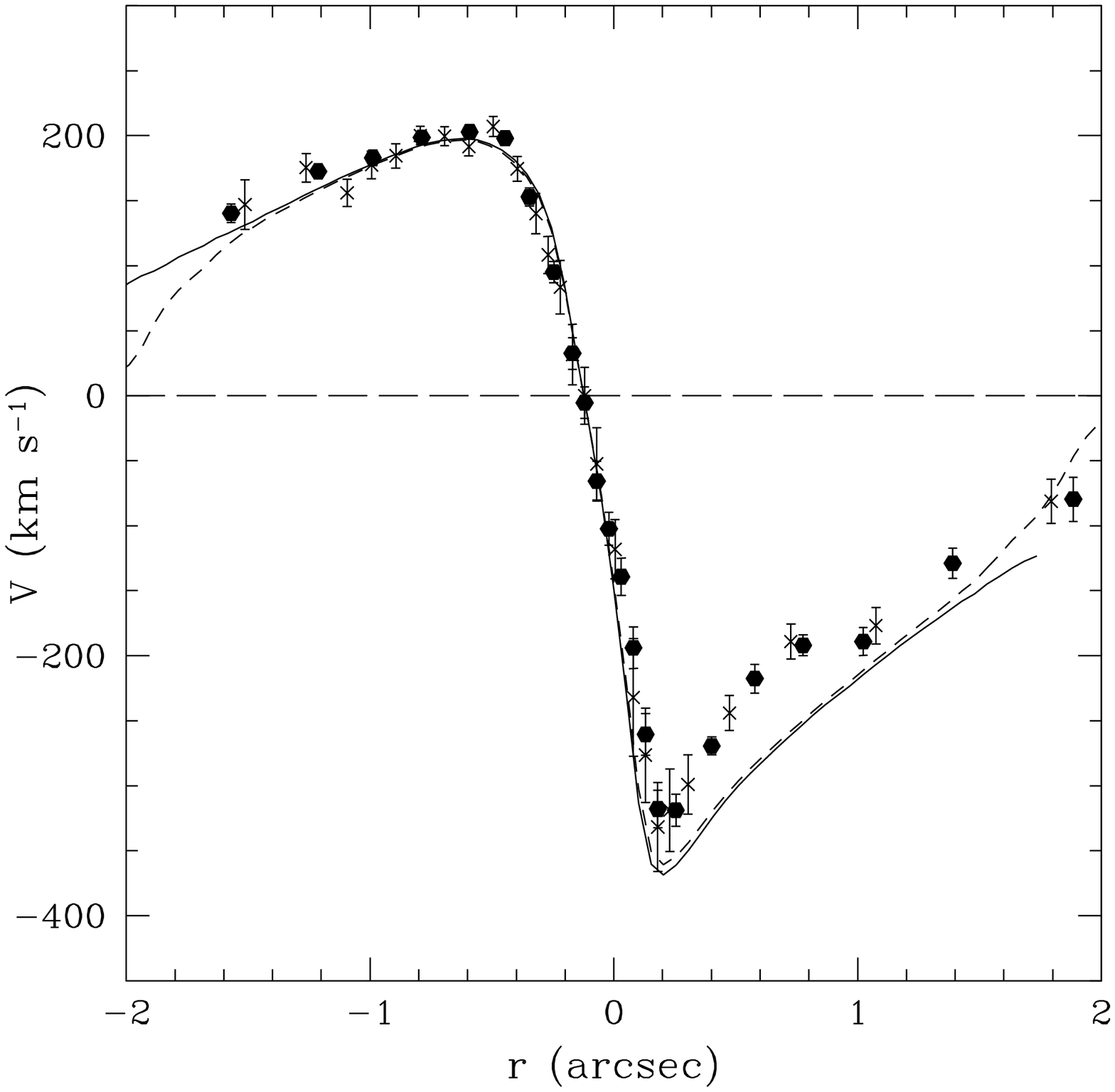}
\caption{Rotation speed $V$ along the axis at PA=39$\arcdeg$, as
observed with STIS resolution (see text). The rotation speed is determined by
fitting the LOSVD to the Gauss-Hermite expansion (\ref{losvd}).  The best
aligned model is shown in the left panel and the best non-aligned model is
shown in the right panel. The solid curve is bulge-subtracted and the dashed
curve includes the bulge. The bulge-subtracted STIS data from \citet{ben03}
are shown as black dots (FCQ) and crosses (FQ).}
\label{fig:rota}
\end{figure}

\begin{figure}
\plottwo{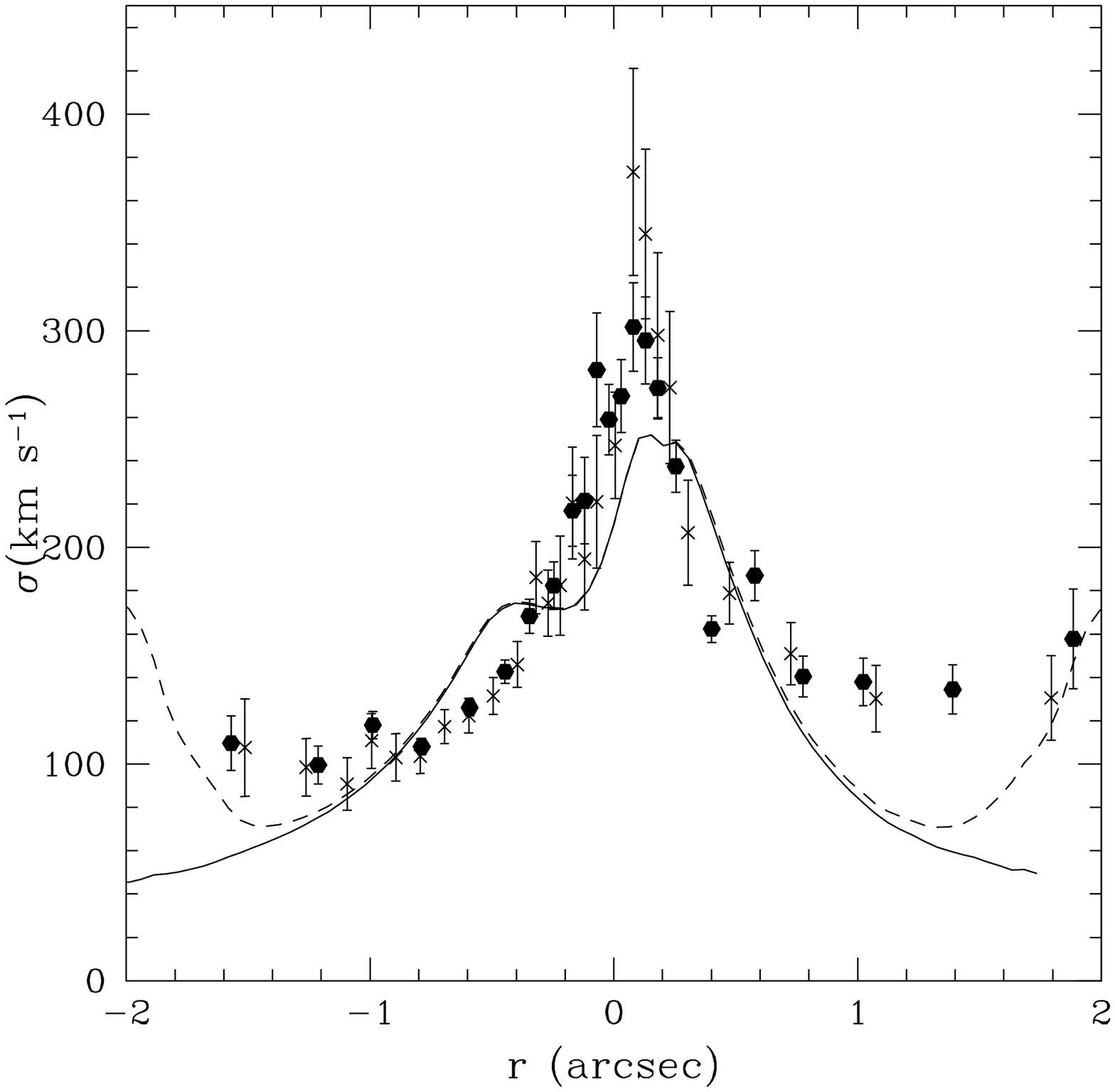}{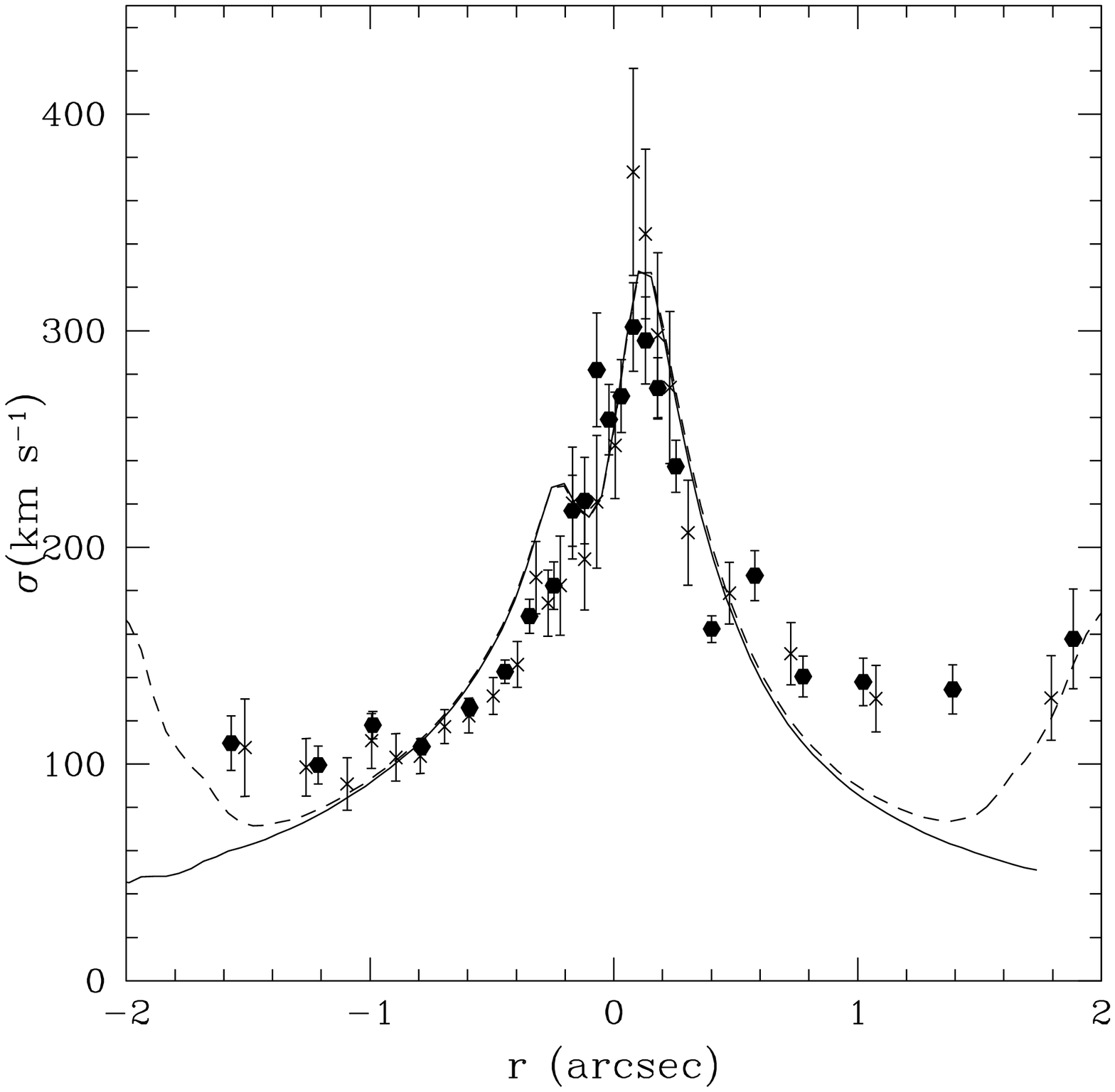}
\caption{Velocity dispersion $\sigma$ along the axis at
PA=39$\arcdeg$, as observed with STIS resolution (see text),
determined by fitting the LOSVD to the Gauss-Hermite expansion
(\ref{losvd}). The best aligned model is shown in the left panel and the
best non-aligned model is shown in the right panel. The solid curve is
bulge-subtracted and the dashed curve includes the bulge. The bulge-subtracted
STIS data from \citet{ben03} are shown as black dots (FCQ) and crosses (FQ).}
\label{fig:dispa}
\end{figure}

\begin{figure}
\plottwo{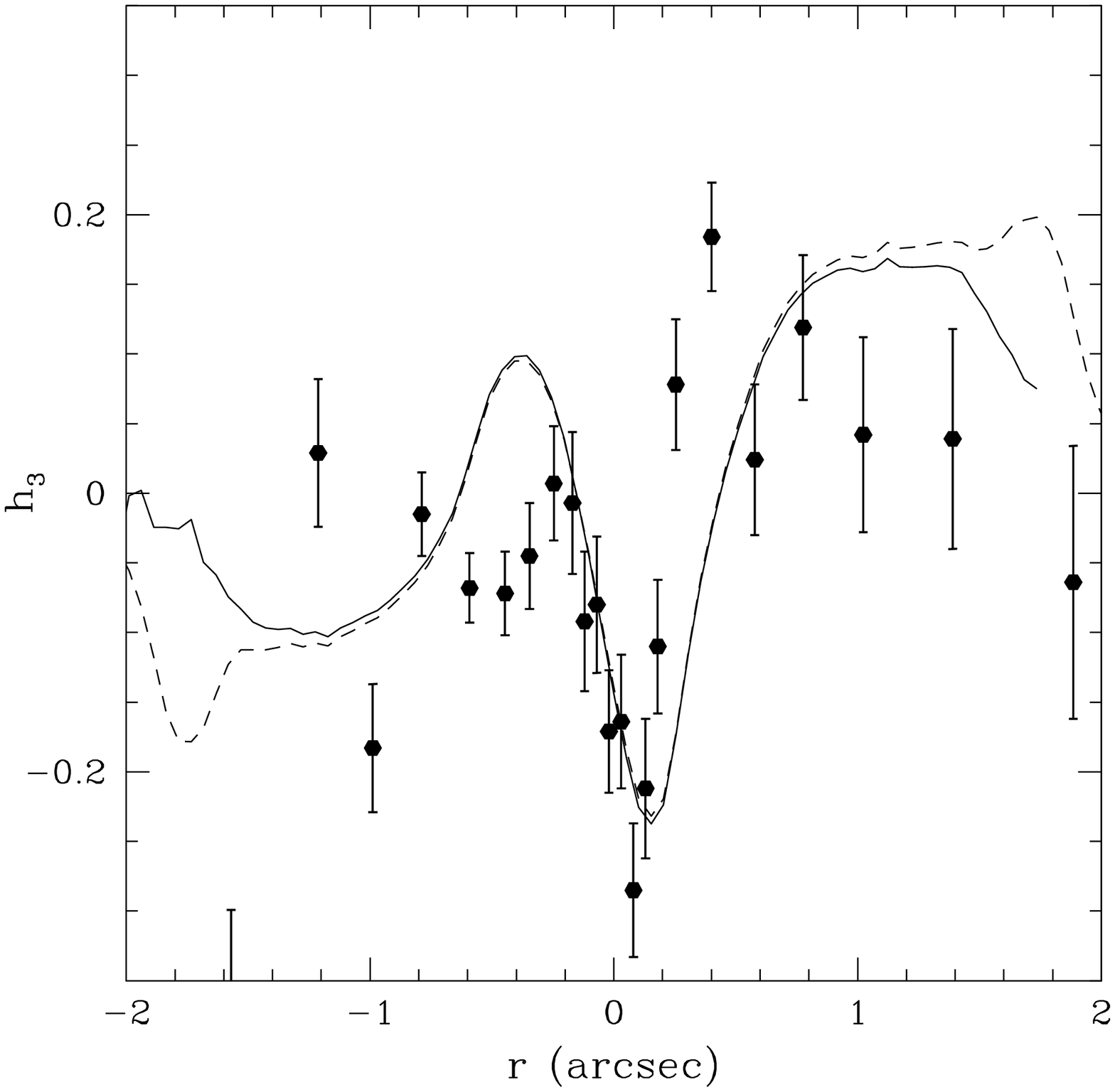}{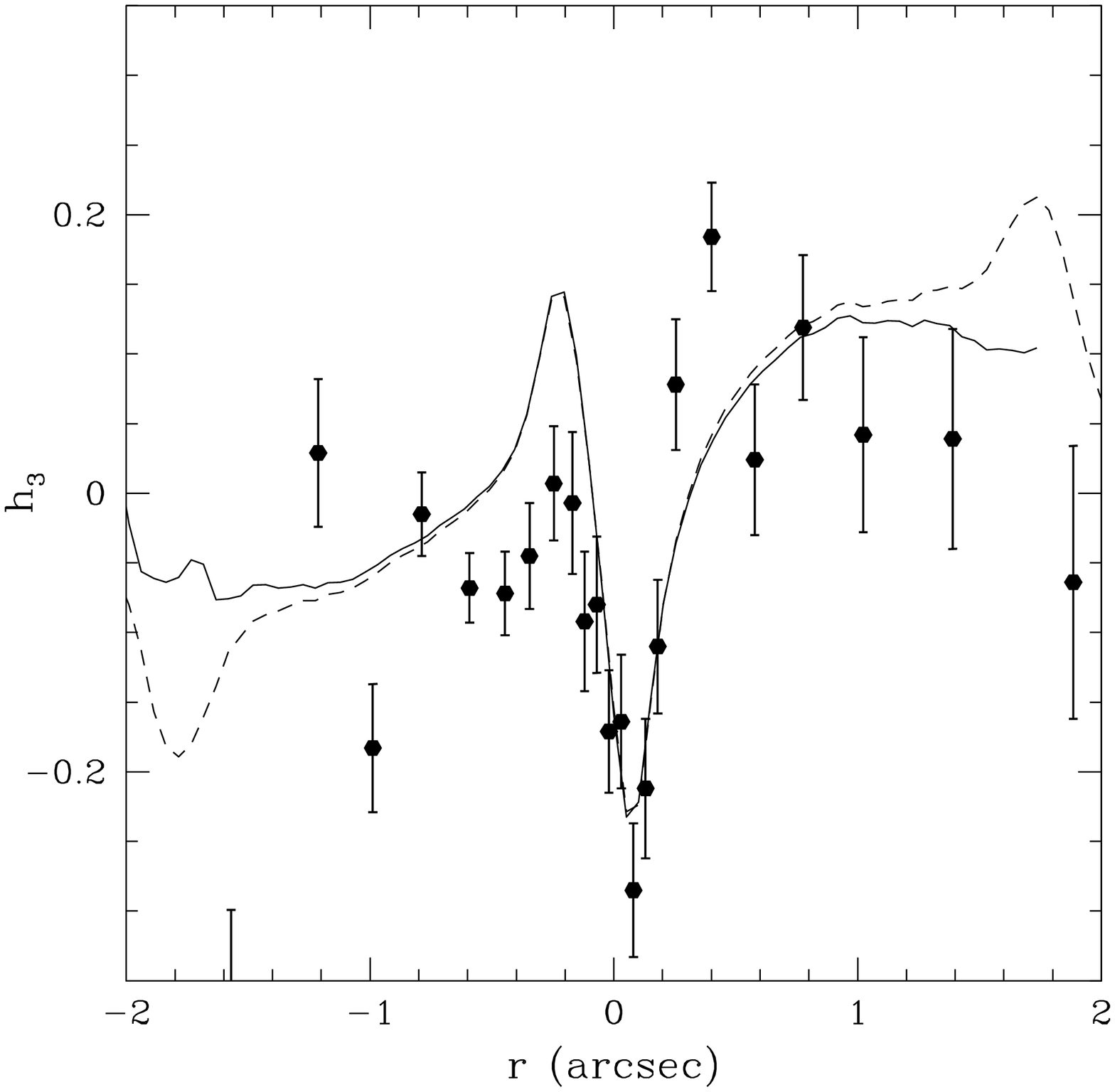}
\caption{The Gauss-Hermite parameter $h_3$ along the axis at
PA=39$\arcdeg$, as observed at STIS resolution. The best aligned model is
shown in the left panel and the best non-aligned model is shown in the right
panel. The solid curve is bulge-subtracted and the dashed curve includes the
bulge. The bulge-subtracted STIS data from \citet{ben03} are shown as black
dots (FCQ).}
\label{fig:hthreea}
\end{figure}

\begin{figure}
\plottwo{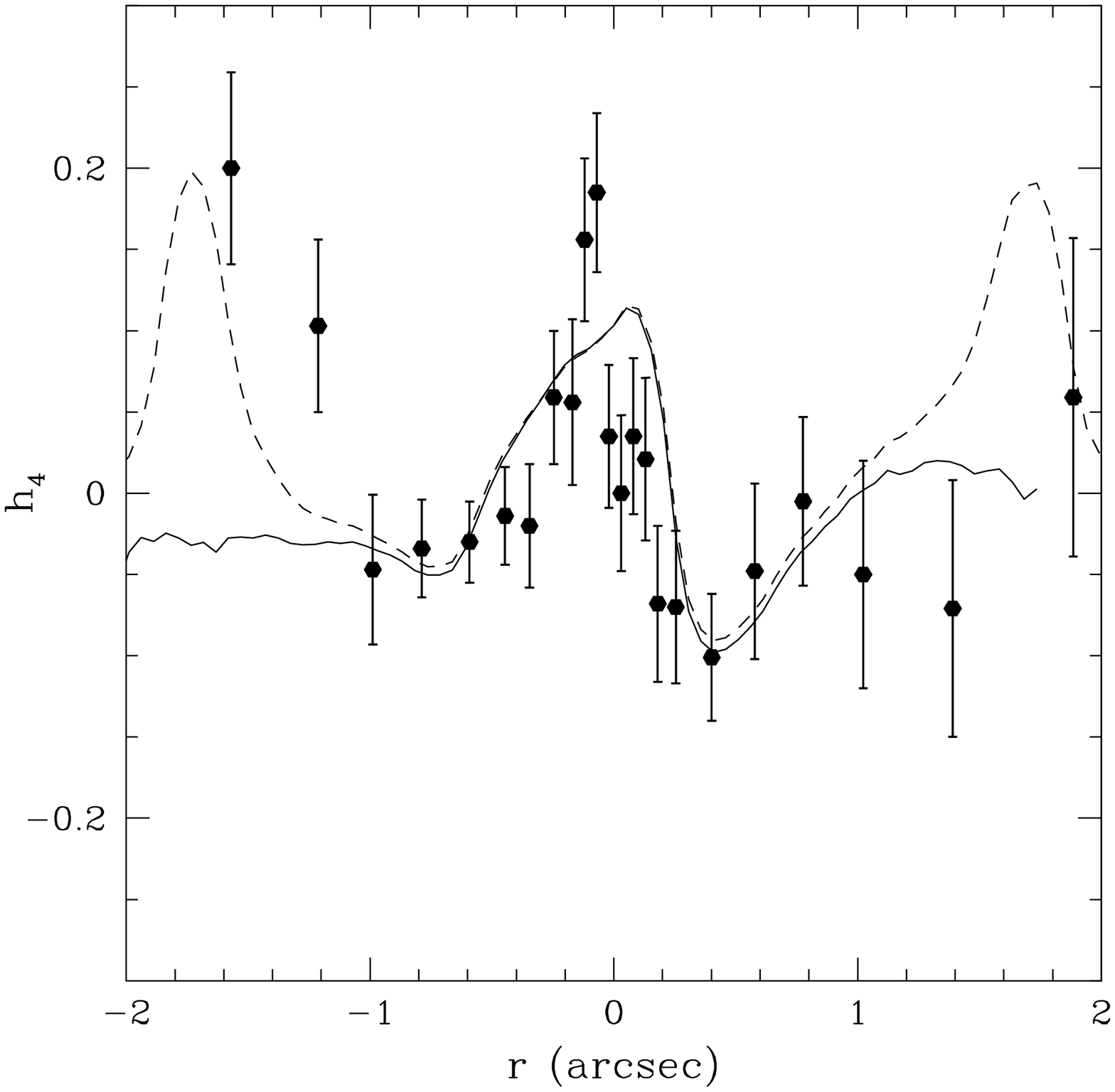}{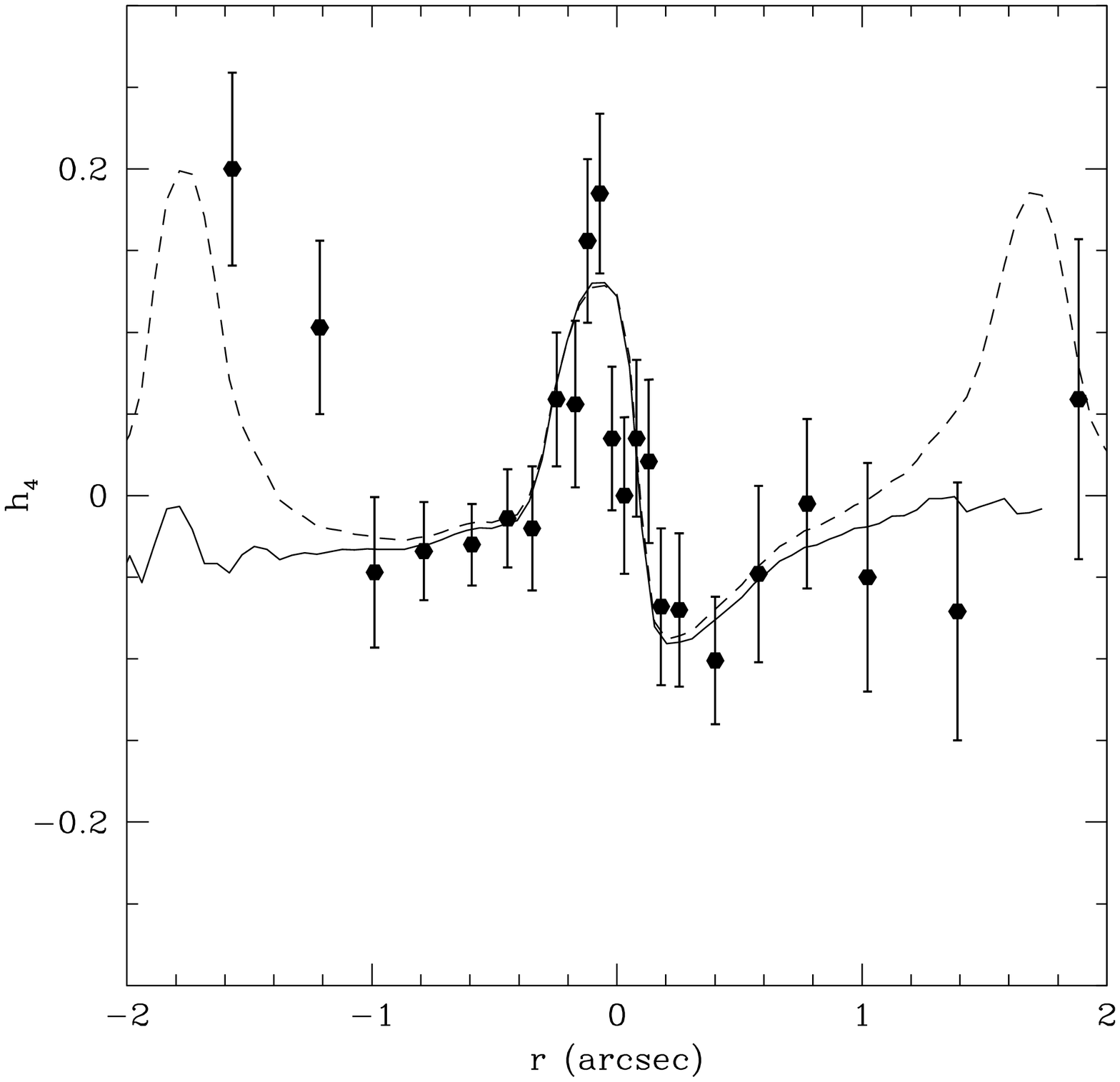}
\caption{The Gauss-Hermite parameter $h_4$ along the axis at
PA=39$\arcdeg$, as observed at STIS resolution. The best aligned model is
shown in the left panel and the best non-aligned model is shown in the right
panel. The solid curve is bulge-subtracted and the dashed curve includes the
bulge. The bulge-subtracted STIS data from \citet{ben03} are shown as black
dots (FCQ).}
\label{fig:hfoura}
\end{figure}

The kinematics of the best-fit aligned and non-aligned models that would be
observed with STIS resolution are shown in Figures
\ref{fig:rota}--\ref{fig:hfoura}. The solid curves are
bulge-subtracted and the dashed curves include the bulge. The data,
shown as dots (FCQ) and crosses (FQ), are obtained at PA=$39\arcdeg$
and are bulge-subtracted. 

In contrast to the KB99 data, the FQ and FCQ results are substantially
different for the STIS data. The difference arises because FQ assumes Gaussian
LOSVDs while FCQ simultaneously derives velocities, dispersions, $h_3$, and
$h_4$ by fitting the LOSVD with a Gauss-Hermite series. Thus, the results from
the two techniques do not agree where LOSVDs have strong wings and the higher
order Gauss-Hermite moments are large. Since we mimic the FCQ procedure in our
calculations, one should compare the model curve with the FCQ results (the
dots).

The non-aligned model is remarkably successful at reproducing the rotation
curve on the P1 side, as well as the rotation gradient through the center
(right panel of Fig.\ \ref{fig:rota}). On the anti-P1 side the peak rotation
speed in the model is too high by 16\% (see Table \ref{tab:kinres}) and the
model rotation curve outside the peak is also too high. 

The displacement of the zero of the rotation curve is 0\,\farcs12 toward P1 in
both the data and the non-aligned model. 

The dispersion peak in the non-aligned model is 9\% too high (Fig.\
\ref{fig:dispa}). The dispersion peak position is reproduced correctly---in
the data the peak is displaced in the anti-P1 direction by 0\,\farcs08,
compared to 0\,\farcs10 in the model. 

At this resolution, both models show a secondary peak in the dispersion curve
(Fig.\ \ref{fig:dispa}). This peak is a consequence of the hole in the middle
of the model surface density distribution. The data do not conclusively favor
or disfavor a feature at this location. One can possibly make this feature
less prominent by making the central surface-brightness gradient
shallower. However, we have been unable to remove the central hole completely
without drastically worsening the fit to the photometry. Thus, we believe that
a secondary peak in the dispersion profile should generally be present at some
level in the type of model proposed here.

The model dispersions are much lower than the dispersions in the data at radii
around $1\arcsec$. We believe that this discrepancy arises because the KB99
bulge model and our disk model do not have enough flexibility to match the
photometry and kinematics in this region; however, this problem should not
affect our models at smaller radii. 

Both the aligned and non-aligned models are quite successful in reproducing
the overall shape of the profiles of the Gauss-Hermite parameters $h_3$ and
$h_4$ (Figs.\ \ref{fig:hthreea} and \ref{fig:hfoura}). In particular, the
models reproduce the shape, amplitude, and location of the sharp dip in $h_3$
and the peak in $h_4$. The agreement between data and models is particularly
impressive since the model is derived by fitting only to the low-resolution
mean velocity and dispersion profiles.

\begin{figure}
\plotone{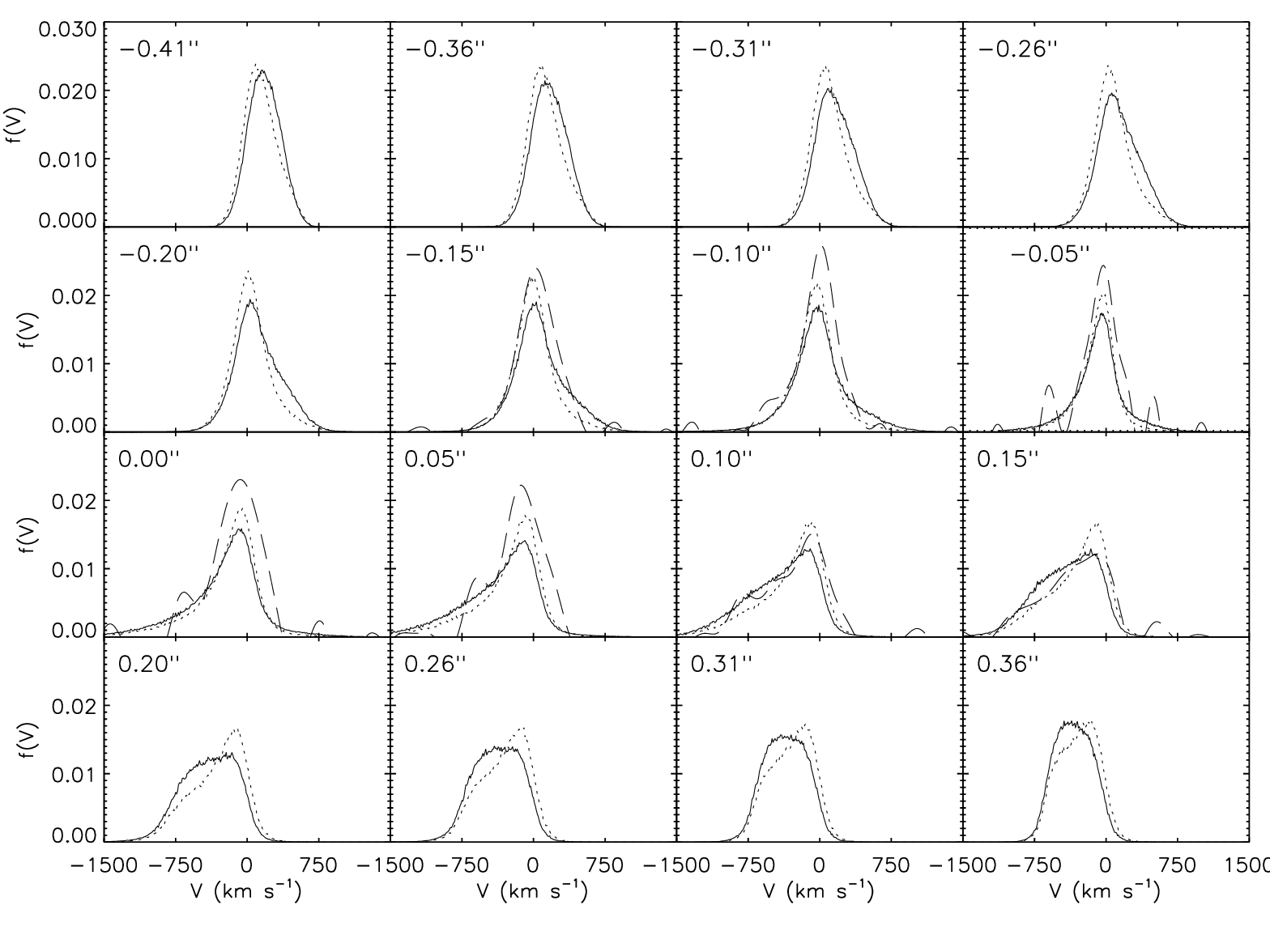}
\caption{The LOSVD distributions near P2 at STIS resolution for the
aligned (dotted) and non-aligned (solid) models. The LOSVDs from the STIS data
are plotted in long dashes (in places the STIS LOSVD is negative and these
sections are not shown). The curves are normalized to unit area. The
velocity zero point is the systemic velocity of M31. Note that not all
the LOSVD bins have the STIS LOSVDs overplotted, because they were
provided only at specific radii, and we could only match some of these
radii to a corresponding pixel on our LOSVD grids.}
\label{fig:losvd}
\end{figure}

Figure \ref{fig:losvd} shows the LOSVDs for each model at a few locations near
P2. These LOSVDs are calculated at STIS resolution along PA=$39\arcdeg$, and
do not include the bulge. Near the BH, the LOSVDs are very asymmetric and have
strong wings on the prograde side of the mean rotation velocity. These
features are also seen in the STIS LOSVDs (plotted here in long
dashes). However the model wings are more extensive than in the data.

Note that the model LOSVDs are constrained to be positive, while the
STIS LOSVDs are not; thus the latter show (unphysical) low-amplitude
oscillations that are not present in the models.

\subsection{Internal dynamics of the disk}

\label{sec:relax}

T95 suggested that two-body relaxation would lead to significant thickening of
a thin disk if the disk age were comparable to the Hubble time. In this
subsection we investigate this process quantitatively for our disk models. For
simplicity, we shall neglect the mean eccentricity of the disk, that is, we
approximate the disk as axisymmetric.

\begin{figure}
\plottwo{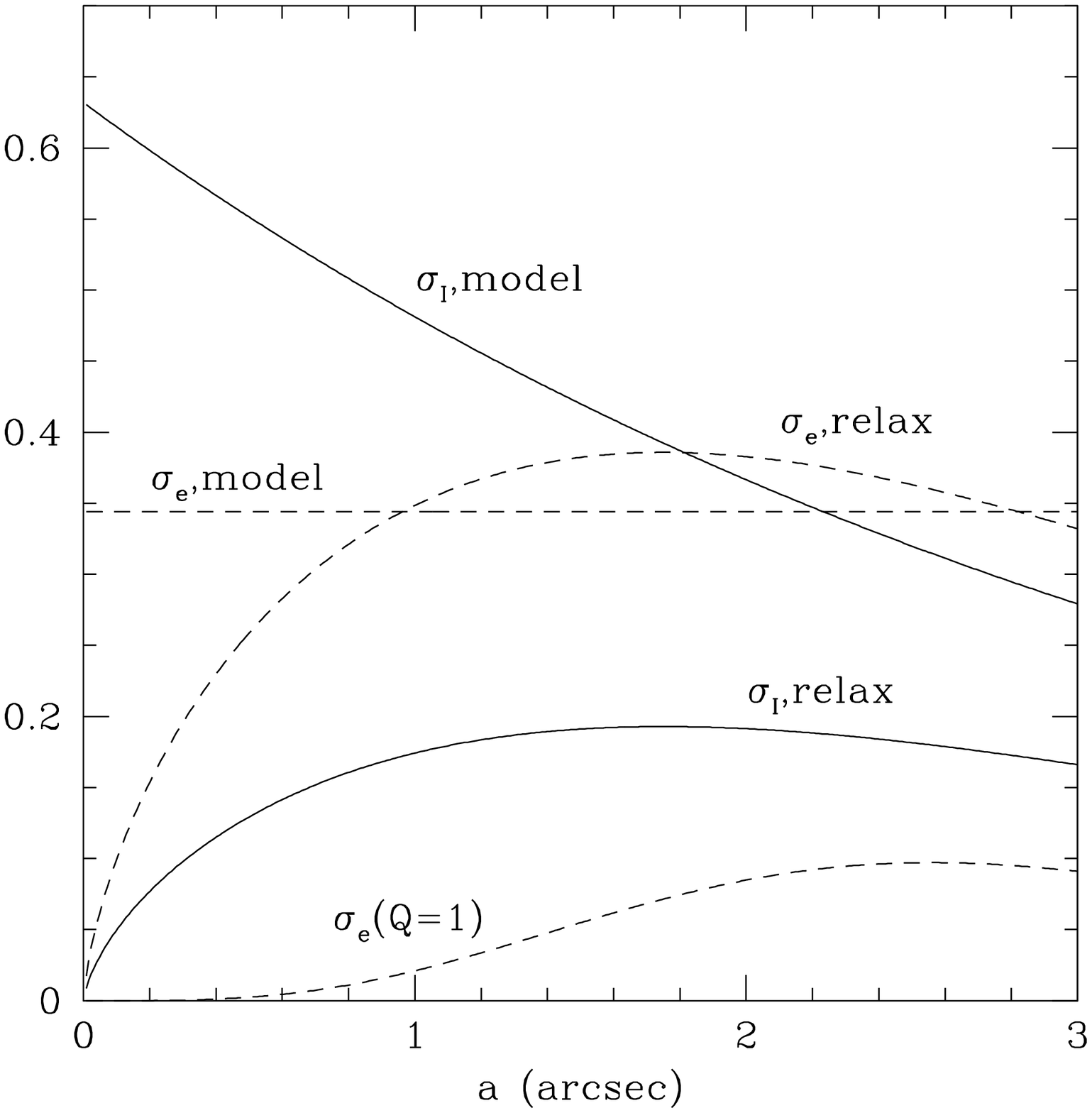}{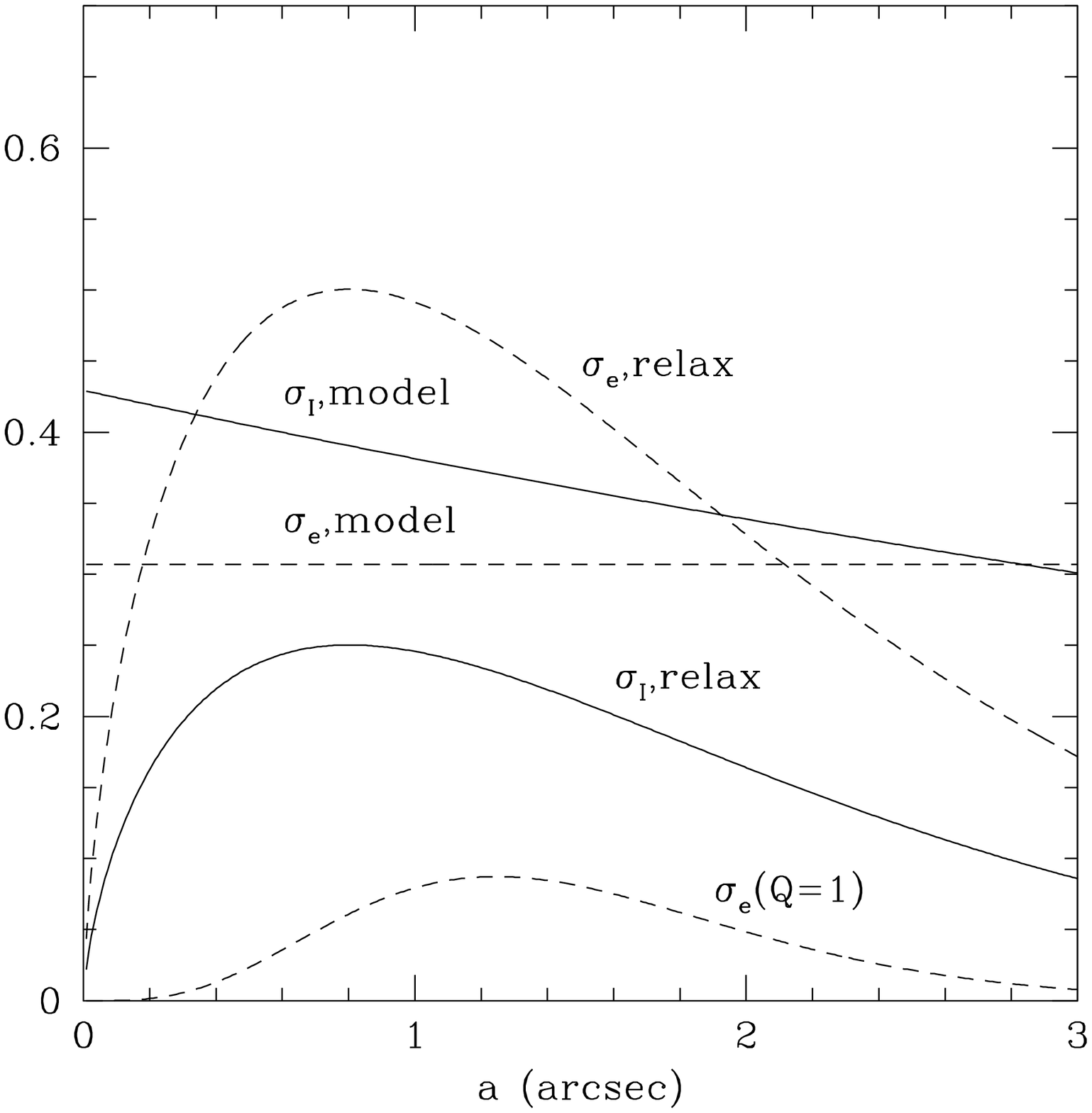}
\caption{The inclination and eccentricity dispersions $\sigma_I$ (solid lines)
and $\sigma_e$ (dashed lines) in the aligned model (left panel) and the
non-aligned model (right panel). The quantities $2\sigma_I^2$ and
$2\sigma_e^2$ are the mean-square inclination and eccentricity at a given
semimajor axis in a disk with mean eccentricity ${\bf e}_m=0$. Also shown are
the dispersions expected from two-body relaxation if the disk age is
$10^{10}\yr$, from equation (\ref{eq:relax}) and the relation
$\sigma_I=0.5\sigma_e$. The curve labeled $\sigma_e(Q=1)$ is the minimum
eccentricity dispersion required for axisymmetric gravitational stability 
(eq.\ \ref{eq:toomre}).}
\label{fig:relax}
\end{figure}

Axisymmetric gravitational stability in a razor-thin disk requires that
Toomre's $Q$ parameter exceeds unity, where \citep{too64,bt87}
\be
Q\equiv 0.30\sigma_e{M_\bullet\over\Sigma(a)a^2};
\label{eq:toomre}
\ee
here $\Sigma(a)$ is the surface density at radius $a$ and $\sigma_e$ is the
rms value of one component of the eccentricity vector or $2^{-1/2}$ of the rms
scalar eccentricity. Using the surface-density distributions in the aligned
and non-aligned models, we find that the value of $\sigma_e$ required for
stability never exceeds 0.1 (Fig.\ \ref{fig:relax}); once the non-zero disk
thickness is included, the required value would be even lower. Thus the
aligned and non-aligned disk models, which have $\sigma_e=0.34$ and $0.31$,
are safely stable.

Let us assume that the disk formed $t_0=1\times10^{10}\yr$ ago, and was
initially cold (i.e. $\sigma_e$ just large enough for gravitational
stability). Two-body relaxation will gradually increase both the rms
eccentricity and the rms inclination of the disk stars. This process of ``disk
heating'' or ``viscous stirring'' has been studied thoroughly in
protoplanetary disks. It is found
that disk heating leads to a specific ratio between the rms inclination and
eccentricity, $\sigma_I\simeq0.5\sigma_e$, and that \citep{osi02}
\be
\langle I^2 \rangle^2(t_0)= 4.52\Omega t_0
{\Sigma(a) a^2\over M_\bullet}{m_*\over M_\bullet}\ln \Lambda, \qquad
\Lambda=0.6{M_\bullet\over m_*}\langle I^2\rangle^{3/2};
\label{eq:relax}
\ee
where $\langle I^2\rangle=2\sigma_I^2$ is the mean-square inclination,
$\Omega(a)=(GM_\bullet/a^3)^{1/2}$, and $m_*$ is the stellar mass. In deriving
this formula we have neglected the time variation of $\langle I^2\rangle$ in
the factor $\Lambda$, since it only appears in a logarithm, and we have
assumed that the disk is in the dispersion-dominated regime, where $\langle
I^2\rangle^{1/2} \gg (m_*/M_\bullet)^{1/3}\simeq 0.002$.

The predicted eccentricity and inclination dispersions, $\sigma_I=\langle
I^2\rangle^{1/2}/\surd{2}$ and $\sigma_e=2\sigma_I$, are shown in Figure
\ref{fig:relax} 
for the aligned and non-aligned models. We have shown both the theoretical
predictions derived from equation (\ref{eq:relax}) and the surface density and
BH mass from Table \ref{tab:parm}, and our empirical fits to the eccentricity
and inclination dispersions, also from Table \ref{tab:parm}. Note that the
theoretical predictions should be robust, since the predicted value of
$\sigma_I$ depends on only the fourth root of the age, surface density,
stellar mass, etc. (eq. \ref{eq:relax}); the principal uncertainty is probably
our assumption that the disk is axisymmetric.

Two-body relaxation requires that the disk axis ratio in the radius range
0.5--$1\arcsec$ is at least 0.13--0.17 in the aligned model, and 0.23--0.25 in
the non-aligned model (these values are $\sigma_I$, which equals the ratio
$\langle z^2\rangle^{1/2}/r$ in a circular disk). In the aligned model, the
thickness required by relaxation is much smaller than the model thickness
($\sigma_I=0.58$--0.55 in the same radius range), and thinner disks are
strongly excluded by the photometry. Thus, if this model is correct, some
process other than two-body relaxation must determine the thickness. The
non-aligned model is thinner, $\sigma_I=0.38$--0.41, and the predicted
thickness due to relaxation is larger, so it is possible that in this case the
thickness is largely determined by two-body relaxation. In aligned disks there
is good agreement between the eccentricity dispersion $\sigma_e$ due to
relaxation and the best-fit value in the model, while in the non-aligned model
the eccentricity dispersion due to relaxation actually exceed the best-fit
value in the model. We suspect that these disagreements are consistent with the
uncertainties in the model.

\section{Discussion} \label{discuss}

The extremely short dynamical time ($1.5\pc/200\kms\simeq 7500 \yr$) of the
M31 nucleus strongly suggests that it is in dynamical equilibrium, an
assumption that we adopt throughout this paper. The M31 nucleus reveals
by far the richest phenomenology of any equilibrium stellar system, and hence
provides a unique challenge for stellar dynamics. 

We have fitted parametrized eccentric-disk models of collisionless stellar
systems to the photometry and kinematics of the nucleus. These models explain
most of the features in the nucleus successfully. However, despite the large
number (11--13) of free parameters and our efforts to use plausible fitting
functions for the DF, there remain significant differences between our
best-fit models and the observational data: for example, we are unable to fit
precisely the separation and relative surface brightnesses of the P1 and P2
components (Fig.\ \ref{fig:slit}). The important question of whether these
discrepancies arise from shortcomings of the eccentric-disk hypothesis,
approximations in our model (e.g. the orbits are assumed to be Keplerian), the
limited number of bright stars in the nucleus, or the limited flexibility of
our parametrization remains to be determined.

The best-fit black-hole mass is $1.0\times 10^8M_\odot$ (Table
\ref{tab:parm}); this estimate is likely to be about 10\%--20\% too high
because we have neglected the contribution of the nuclear disk stars to the
total mass [the ratio of the disk mass inside $1\arcsec$ to the BH mass is
$\sim0.15$ (Table \ref{tab:photres})]. Previous estimates of the BH mass
include 3--$7\times10^7M_\odot$ \citep{dr88}, $10^{6.5}$--$10^8M_\odot$
\citep{kor88}, 4--$5\times10^7M_\odot$ \citep{rbd90}, 
$7\times10^7M_\odot$ \citep{bac94}, $7.5\times10^7M_\odot$ (T95),
7--$10\times10^7M_\odot$ \citep{ec97}, $(3\pm1.5)\times10^7M_\odot$ (KB99),
and $(7{+1.5\atop-3.5})\times10^7M_\odot$ (B01). The relatively small mass
preferred by KB99 is found by a quite different method than the other
estimates (they use the displacement between P2B and the center of the bulge,
as opposed to the rotation curve and velocity dispersion profile), and may be
subject to systematic errors despite the great care taken by KB99 to measure
this small displacement accurately. For comparison, the correlation between BH
mass and bulge dispersion observed in other nearby galaxies \citep{tre02}
predicts a mass of $6\times10^7M_\odot$ with an uncertainty of a factor of
two.

Although the models presented in this paper represent a substantial advance in
accuracy and realism, they still have many shortcomings.

\begin{itemize}

\item We assume that the stars in the nuclear disk orbit in the point-mass
potential of the BH, that is, we neglect the gravitational forces from the
disk and bulge. Neglecting the bulge is reasonably safe: within $1\arcsec$ the
mass of the bulge is $<10^6M_\odot$, less than 1\% of the mass of the
BH. Neglecting the disk is a more serious approximation: in our best-fit
models, the disk mass within $1\arcsec$ is $\sim 15$\% of the BH mass (Tables
\ref{tab:parm} and \ref{tab:photres}); thus the effects of the self-gravity of
the disk are small but not negligible. Including the gravitational potential
of the disk will modify the kinematics of the disk orbits, and may
permit new types of orbit family not present in the spherically symmetric BH
potential (e.g. lens orbits and chaotic orbits; see Sridhar \& Touma 1999 and
Sambhus \& Sridhar 2002). An additional consequence of including the disk 
potential is that the eccentric disk will precess, an effect not
included in our model \citep{sam00,sam02}. We believe that properly including
these effects may significantly improve the kinematic fits. However, this
improvement is unlikely to have much effect on the photometry. Thus, in
particular, our conclusion that non-aligned models fit the photometry better
than aligned models is unlikely to change.

\item The eccentric-disk model requires that the apsides of the disk stars
precess uniformly so that the disk maintains its apsidal alignment; this
requirement provides an additional strong constraint on the mass and
eccentricity distribution in the model that we have not exploited.
\citet{sta99} has argued that uniform precession requires that the 
disk have a steeply declining eccentricity gradient and a change in direction
of the eccentricity vector. The steep eccentricity gradient is present in our
models, but is required by the high surface brightness at P1 rather than for
any dynamical reason (eq. \ref{thinratio}). The eccentricity does not reverse
sign to any significant extent (see Fig. \ref{fig:meane}). Statler's argument
for the eccentricity reversal is plausible but suspect, for at least two
reasons: (i) The same argument, applied to a slightly eccentric test-particle
orbit in an axisymmetric disk, would predict that the periapsis would precess
in a prograde direction, but orbits in continuous disks precess in a
retrograde direction ($\dot\varpi=\Omega-\kappa$, where $\Omega$ and $\kappa$
are the azimuthal and radial frequencies, and in general
$\kappa>\Omega$). (ii) \citet{tre01} has shown that self-consistent $m=1$
linear normal modes in nearly Keplerian disks require non-zero velocity
dispersion or thickness, effects that are not included in Statler's numerical
calculations or qualitative discussion.

\item Our models contain only prograde orbits, and retrograde orbits may also
contribute to the structure of the M31 nucleus; \citet{sam02} find that
retrograde orbits significantly improve their fits and this possibility
deserves to be explored further.

\item We have used parametrized models for the DF; thus, we are unable to
determine whether the differences between our models and the data are due to
shortcomings of the model or the parametrization. A more flexible,
but computationally expensive, approach would be to compile an orbit library
and use a penalized maximum-likelihood method such as maximum entropy to
construct the best-fit smooth DF. 

\item The bulge model of KB99 does not match on well to our
disk models, and there are discrepancies in the photometry (Fig.\
\ref{fig:slit}) and the velocity dispersion (Figs.\ \ref{fig:disp} and
\ref{fig:dispa}) at radii $\gtrsim 1\arcsec$. More flexible disk and/or bulge
models are needed to fit this region satisfactorily. 

\item We have not explored the effects of noise due to the limited number of
bright stars in the nucleus. Even in the brightest parts, there are fewer than
$10^3$ stars per $(0\,\farcs1)^2$ area (L98), suggesting that the
uncertainties due to Poisson statistics are at least a few percent in the STIS
kinematics; the errors are likely to be even larger in the photometry because
deconvolution enhances Poisson noise. 

\item We have fit the kinematics only along the P1-P2 axis. Integral-field
spectroscopy by B01 provides the mean velocity and velocity dispersion across
the entire two-dimensional face of the nucleus, although at lower spatial
resolution. We have shown that this data is approximately consistent with our
model (Figures \ref{fig:align2dkin} and \ref{fig:nonalign2dkin}) but it should
be included in the model fits as well.

\end{itemize}

A striking feature of our simulations is that the non-aligned model fits the
data substantially better than the aligned model. As stated above, this
conclusion is unlikely to be an artifact of our neglect of the self-gravity of
the disk, since the improved fit is apparent in both the photometry and the
kinematics. Our best-fit inclination is $54\arcdeg$ compared to an inclination
of $77\arcdeg$ for the large-scale M31 disk. This inclination is consistent
with values derived by other investigators when modeling the nucleus as a thin
disk: $55\arcdeg$ (B01), $50\arcdeg$ \citep{pen02}, and $52\arcdeg$
\citep{sam02}. The position angle of the line of nodes of the disk on the sky
plane in the non-aligned model is $\theta_l+90\arcdeg=47\arcdeg$, which is
reasonably close to the angle derived by B01 ($54\arcdeg$ for the model in
their Fig.\ 20). 

Most of the fitted parameters in the aligned and non-aligned models are rather
similar (Table \ref{tab:parm}).  However, the non-aligned model has a lower
peak in the mean eccentricity $e_m$ (Fig. \ref{fig:meane}); this is turn is
mostly controlled by the parameter $\alpha$ (eq. \ref{eccform}), which is 30\%
smaller in the non-aligned model. In the observations, $\alpha$ primarily
affects the positions and amplitudes of the velocity maxima, while not greatly
influencing the dispersion curve. Decreasing $\alpha$ in the aligned model
significantly improves the fit to the velocity curve. The parameter $a_0$,
which controls the surface-brightness scale length of the disk
(eq. \ref{surfden}), is more than a factor of three larger in the aligned
model than in the non-aligned model. Decreasing $a_0$ in the aligned model
brings more stars near the BH and therefore can dramatically increase the
dispersion peak; this parameter does not greatly affect the rotation
velocity. However, if these changes to $\alpha$ and $a_0$ are made to the
aligned model, the fit to the photometry worsens dramatically, giving a P2
peak that is very much brighter than P1, and not matching the general features
of the two-dimensional photometry. Conversely, if one changes the orientation
angles of the non-aligned model to align the disk with the large-scale disk
of M31, both the photometric and kinematic fits become worse.

Thus, it appears that an aligned eccentric disk can match the photometry or
the kinematics but not both simultaneously. 

There are some arguments that non-aligned nuclear disks may be present in
other galaxies. (i) The axis of the 0.2-pc radius maser disk in the center of
NGC 4258 is $119\degr$ from the axis of its host galaxy \citep{miy95}. (ii)
The S0 galaxy NGC 3706 contains an edge-on disk of radius $\sim 20\pc$ that is
tipped by $\sim30\arcdeg$ to the major axis of the elliptical isophotes at
larger radii \citep{lau02}. (iii) The jets in Seyfert galaxies, which are
presumably perpendicular to the inner parts of the nuclear disk, are not
perpendicular to the large-scale galaxy disk \citep{sk02}. A counter-argument
is that the isophote twists that should be associated with non-aligned disks
do not appear to be seen near the centers of edge-on disk galaxies, although
this argument has not been quantified in a well-defined sample.

The possibility that the nuclear disk is not aligned with the large-scale M31
disk raises several interesting issues. A natural question is whether the
inner bulge of M31 shares the orientation of the nuclear disk or the
large-scale disk. Normally, of course, the bulge of M31 is assumed to be
aligned with the large-scale disk, since bulges and disks appear to be aligned
in edge-on spiral galaxies. A bulge aligned with the large-scale disk must be
triaxial or barred \citep{sta77,sb94,bl02}. \citet{rui76} has investigated the
alternative possibility that the bulge is axisymmetric but not aligned with the
large-scale disk; in this case she finds that fitting the photometric and
kinematic data requires an inclination of $\sim 55\arcdeg$. The close
agreement of this inclination with the derived inclination of the nuclear disk
is striking but inconclusive.

If the nuclear disk is not aligned with the inner bulge, then it is subject to
dynamical friction from the bulge, which usually damps its inclination
relative to the bulge. Dynamical friction arises because of precession of
the line of nodes of the nuclear disk on the equatorial plane of the
bulge. Numerical simulations of the damping of large-scale warps in
non-spherical halos indicate that the damping time is typically not much
longer than the precession time \citep{dk95,nt95}. The precession time of
the nuclear disk in the potential of the bulge is $\sim 10^7\yr$, unless the
bulge is very accurately spherical (the axis ratio of the isophotes of the
inner bulge is 0.8--0.9; see Kent 1989 and Peng 2002). Thus we expect that a
long-lived nuclear disk should be aligned with the bulge. A possible 
way to evade this conclusion would be to argue that dynamical interactions
with the bulge excite, rather than damp, the inclination of the nuclear disk
\citep{nt95}; one attraction of this approach is that one might
also find that interactions with the bulge excite the mean eccentricity of
the disk. 

\begin{figure}
\plottwo{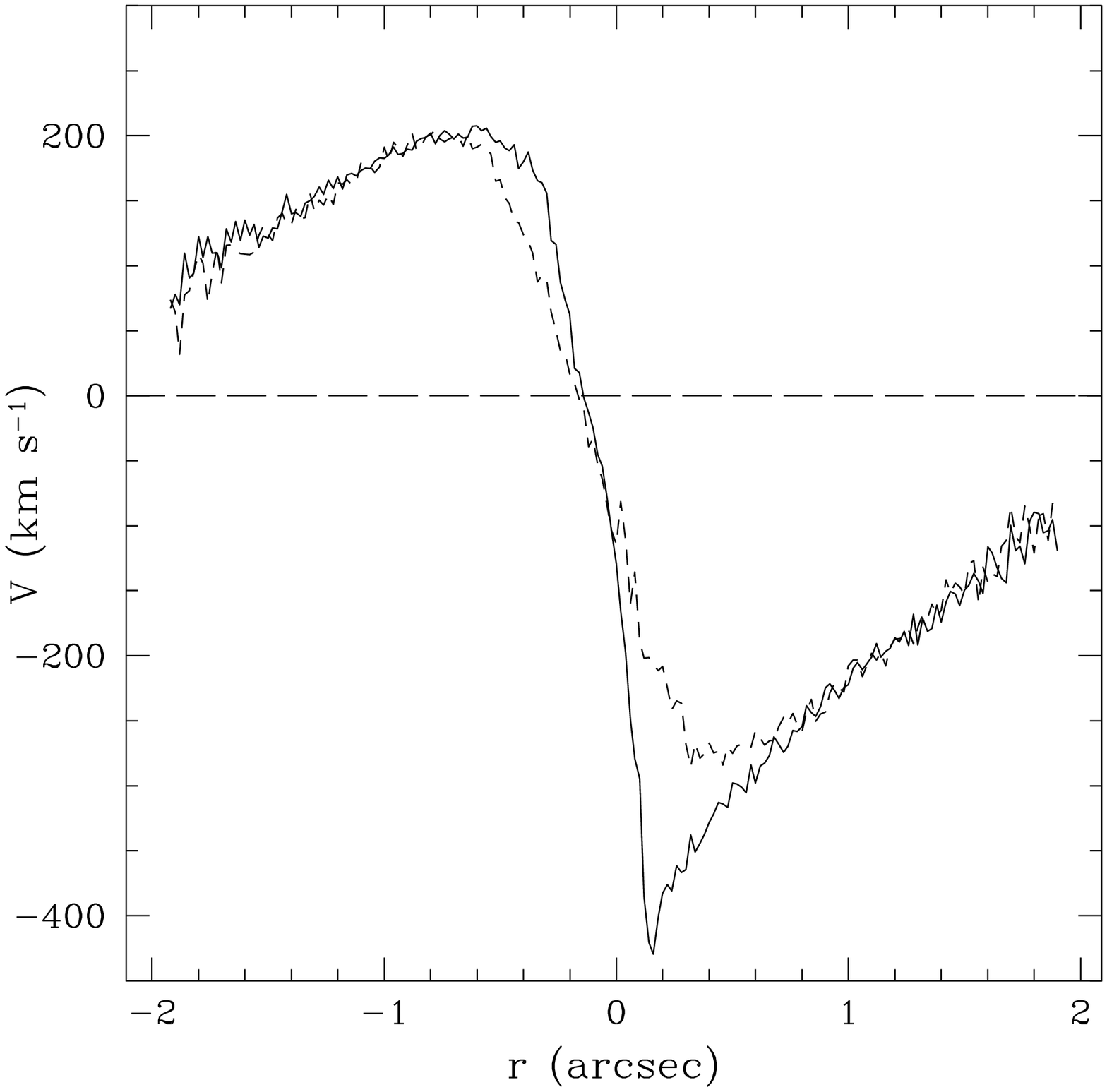}{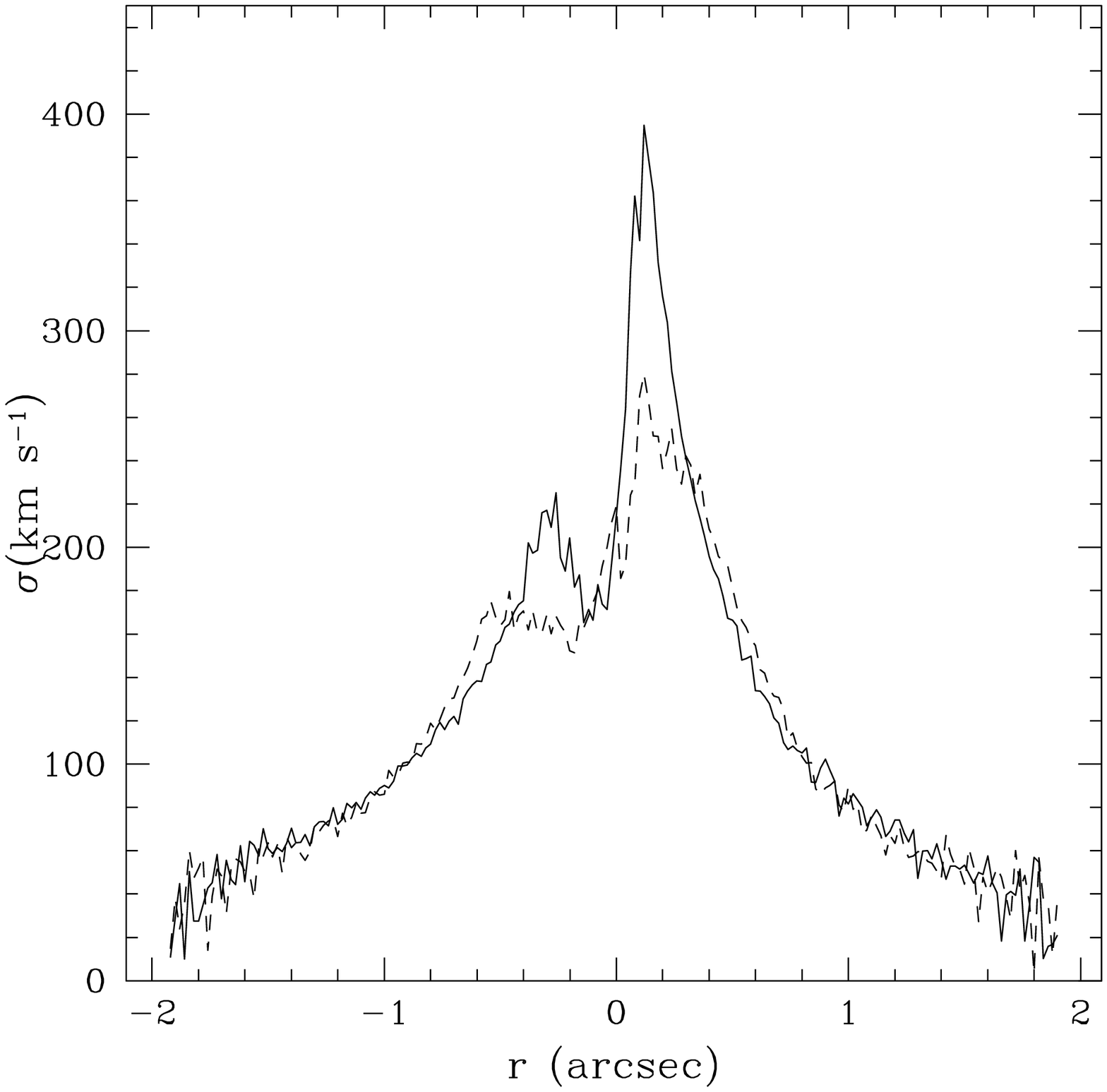}
\caption{The bulge-subtracted rotation speed (left) and velocity dispersion
(right) as defined by the Gauss-Hermite expansion (\ref{losvd}), as observed
along a slit of width 0\farcs02 at PA=39$\arcdeg$. The non-aligned model is
shown using a solid line and the aligned model using a dashed line. The curves
are jagged because of Poisson noise in our Monte-Carlo simulation, which is
smaller than the expected Poisson noise from the limited number of stars in
the actual nucleus. }
\label{fig:vhires}
\end{figure} 

\begin{figure}
\plottwo{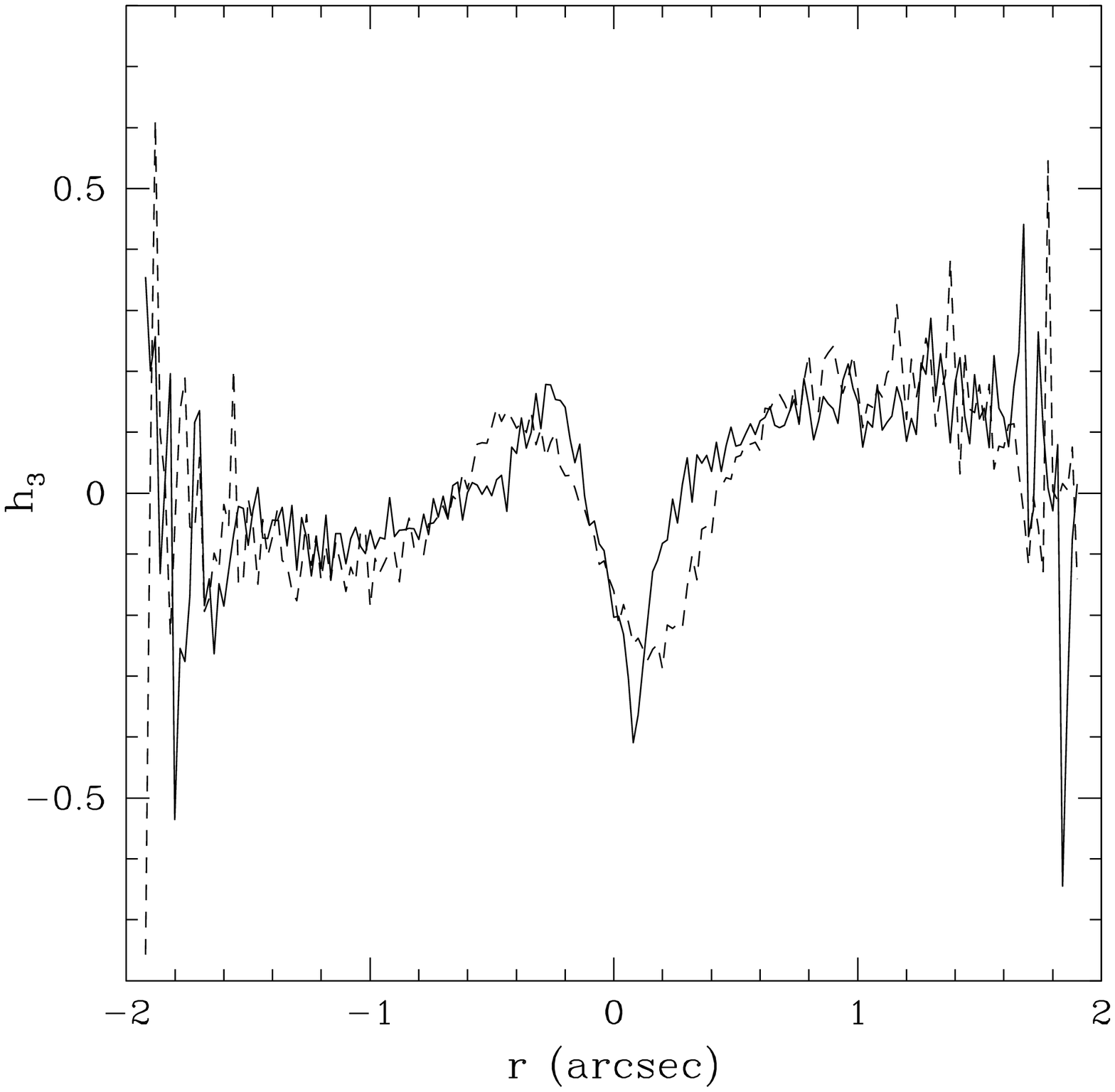}{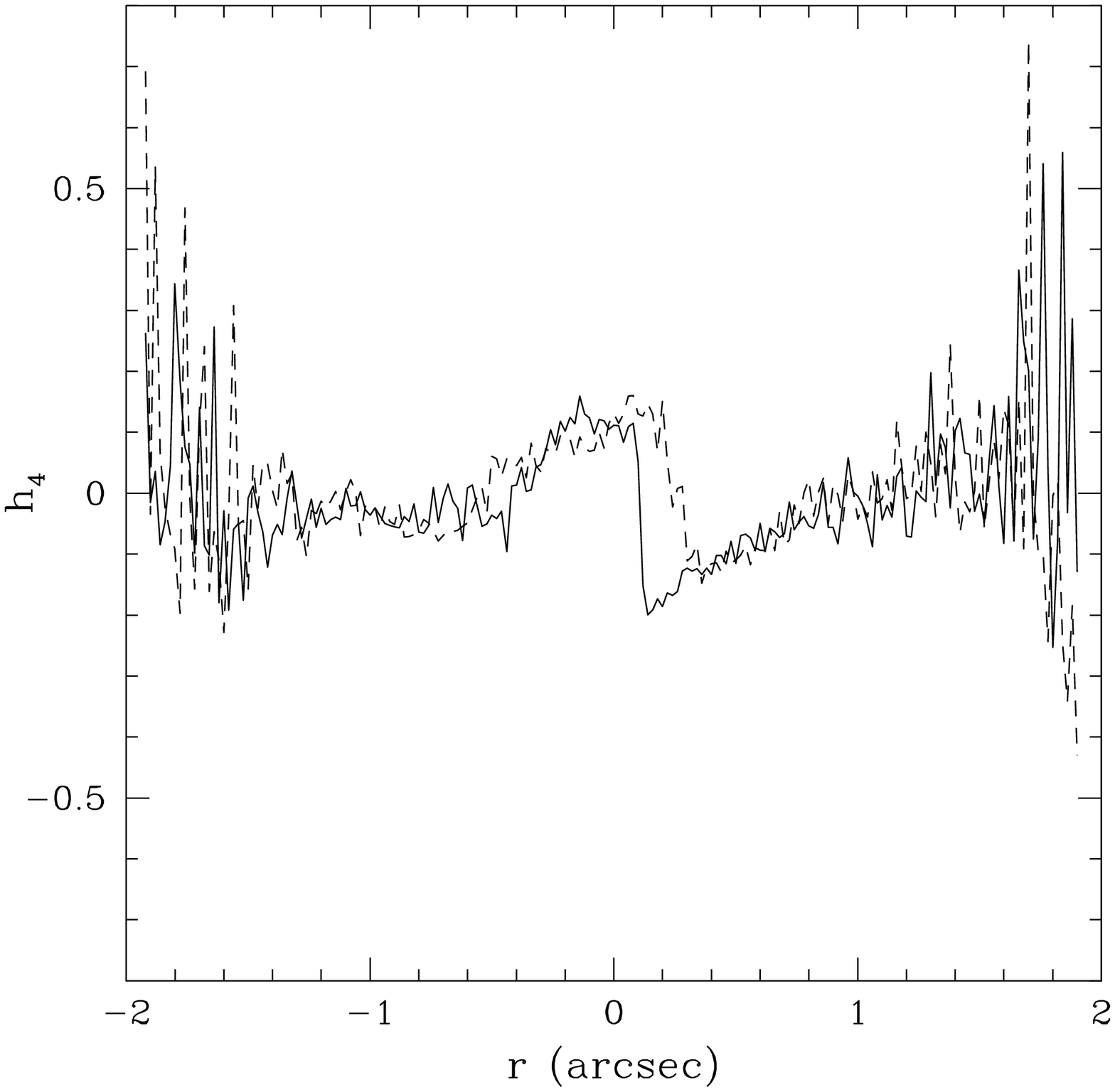}
\caption{The Gauss-Hermite parameters $h_3$ (left panel) and $h_4$ (right
panel), as observed along a slit of width 0\farcs02 at PA=39$\arcdeg$. The
non-aligned model is shown using a solid line and the aligned model using a
dashed line.}
\label{fig:hhires}
\end{figure} 

The current photometric and kinematic data on the nucleus of M31 are
sufficiently accurate and feature-rich to justify the development of
more accurate and flexible dynamical models that incorporate the
improvements listed at the beginning of this Section. Such models
should yield new insights into the formation and structure of
eccentric stellar disks. In the more distant future, we may look
forward to high-resolution imaging of the nucleus using adaptive
optics, interferometric imaging by the Space Interferometry Mission,
and measurements of proper motions in the nucleus, by HST or its
successors or by SIM. In this spirit, Figures \ref{fig:vhires} and
\ref{fig:hhires} show the kinematic parameters of our models as viewed
at a resolution of 0\farcs02. 
\acknowledgements

We thank Ralf Bender, Karl Gebhardt, John Kormendy and Tod Lauer for expert
advice, S.~Sridhar for thoughtful discussions, and the referee, Eric Emsellem,
for many suggestions that improved the presentation and content of the
paper. We also thank Bender and his collaborators for providing their
STIS observations of the M31 nucleus in advance of publication. This
research was supported in part by NSF grant AST-9900316 and by NASA grant
HST-AR-09513.01-A.


\end{document}